\documentclass[useAMS,usenatbib]{mn2e}

\usepackage{graphicx}
\usepackage[T1]{fontenc}
\usepackage{amssymb}
\usepackage{color}
\usepackage{amsmath}
\usepackage{hyperref}

\newcommand{\noun}[1]{\textsc{#1}}
\newcommand{\simapp}{\mathord{\sim}}

\newcommand\ion[2]{#1$\;${\small \uppercase\expandafter{\romannumeral #2}}\relax}%
\newcommand\nodata{ ~$\cdots$~ }%

\title[Photometric Behavior of AM CVn Systems]{Long-term Photometric Behavior of Outbursting AM CVn Systems}

\author[D. Levitan et al.]{David Levitan$^1$\thanks{E-mail: \href{mailto:dlevitan@astro.caltech.edu}{dlevitan@astro.caltech.edu}}, 
Paul J. Groot$^{1,2}$,
Thomas A. Prince$^1$,
Shrinivas R. Kulkarni$^1$,\newauthor
Russ Laher$^3$,
Eran O. Ofek$^4$,
Branimir Sesar$^1$,
and Jason Surace$^3$.
\\ \\
$^1$ Division of Physics, Mathematics, and Astronomy, California Institute of Technology, Pasadena, CA 91125, USA.\\
$^2$ Department of Astrophysics/IMAPP, Radboud University Nijmegen, PO Box 9010, NL-6500 GL Nijmegen, the Netherlands.\\
$^3$ Spitzer Science Center, MS 314-6, California Institute of Technology, Pasadena, CA 91125, USA.\\
$^4$ Benoziyo Center for Astrophysics, Weizmann Institute of Science, 76100 Rehovot, Israel.
}

\begin{document}

\date{Accepted ...  Received ...; in original form ...}

\pagerange{\pageref{firstpage}--\pageref{lastpage}} \pubyear{2014}

\maketitle

\label{firstpage}

\begin{abstract}
The AM CVn systems are a class of He-rich, post-period minimum, semi-detached, ultra-compact binaries. 
Their long-term light curves have been poorly understood due to the few systems known and the long (hundreds of days) recurrence
times between outbursts. We present combined photometric light curves from the LINEAR, CRTS, and PTF synoptic surveys to study
the photometric variability of these systems over an almost 10\,yr period.
These light curves provide a much clearer picture of the outburst phenomena that these systems undergo.
We characterize the photometric behavior of most known outbursting AM CVn systems and establish a relation between their outburst
properties and the systems' orbital periods. We also explore why some systems have only shown a single outburst so far and expand
the previously accepted phenomenological states of AM CVn systems. We conclude that the outbursts of these
systems show evolution with respect to the orbital period, which can likely be attributed to the decreasing mass transfer rate
with increasing period. Finally, we consider the number of AM CVn systems that should be present in modeled synoptic surveys.
\end{abstract}


\section{Introduction}

The AM CVn systems are a rare class of ultra-compact, post-period minimum, stellar binaries with some of the smallest orbital separations known. Ranging
in orbital period from 5 to 65 minutes, they are believed to be composed of a white dwarf
accreting from a lower mass white dwarf or semi-degenerate helium star donor \citep{1967AcA....17..287P,Faulkner:1972cr}.
We refer the reader to \citet{2005ASPC..330...27N} and \citet{2010PASP..122.1133S} for reviews.

As a result of their mass-transferring nature, most AM CVn systems show inherent photometric variability on multiple time-scales,
believed to be largely dependent on the orbital period and mass transfer rate of the particular system. AM CVn system phenomenological
behavior has been separated into two states --- a ``high'' state
corresponding to high rates of mass transfer resulting in an optically thick
  accretion disc around the primary --- and a ``quiescent'' state corresponding to low rates of mass
transfer and an optically thin disc. The high state is generally associated with those systems having orbital periods $< 20\,$min and
the quiescent state with those having orbital periods $> 40\,$min. High state systems exhibit superhump behavior like that found
in some cataclysmic variables \citep[CVs; ][]{WARNER1995}
with photometric variability close to the orbital time-scale at an amplitude of $\mathord{\approx}0.1\,$mag \citep[e.g., ][]{Patterson:2002ys}.

Systems with orbital periods between $\mathord{\approx}20\,$min and $\mathord{\approx}40\,$min 
have been observed to alternate between the high and quiescent states and have behavior similar to that of
super-outbursts in dwarf novae and are thus called ``outbursting'' AM CVn systems \citep[e.g., ][]{2012MNRAS.419.2836R}.
In outburst, these systems are typically 3--5\,mag brighter than in quiescence and these outbursts have been observed
to recur on time-scales from $\mathord{\sim}40\,$d to several years.
Some systems, particularly those at the short-period end, are also observed to have shorter, ``normal'' outbursts that last 1--1.5\,d
and are typically seen 3--4 times between the longer ``super''-outbursts \citep[e.g., ][]{2000MNRAS.315..140K,2011ApJ...739...68L}.
Given the much longer cadences for the data presented here, we are interested only in super-outbursts and will refer to them as just
outbursts, unless explicitly specified.

One of the outstanding questions about AM CVn systems is the disagreement between population density estimates derived
from population synthesis modeling and those calculated from the number of observed systems (see, e.g., \citealt{Carter:2013fk}
for the latest overview).
The intrinsic low luminosity of the systems means few systems have been discovered; the known sample remained under a
dozen for almost 40 years until the availability of the Sloan Digital Sky Survey (SDSS). This also makes obtaining a
systematically identified sample of AM CVn systems large enough to measure the population density difficult. 
The recent availability of large-area surveys has allowed for the identification of AM CVn systems
both from their spectra (or colours) and their aforementioned light curves in a systematic fashion, with relatively well-understood
selection biases. This has led to the number of known AM CVn systems tripling in the last decade and the identification  of
two complementary, systematically-selected sets of systems.

Searches of the SDSS spectroscopic database for He-rich, H-poor sources
have been particularly successful, with nine new systems identified
\citep{2005MNRAS.361..487R,2005AJ....130.2230A,2008AJ....135.2108A,2014MNRAS.439.2848C}.
\citet{2007MNRAS.382..685R} found that the spectroscopic completeness of the SDSS database
in the relatively sparse region of colour-colour space that AM CVn systems are believed to occupy,
and at the faint apparent magnitudes where most systems are expected to be found, was only $\mathord{\sim}20\%$.
A subsequent effort using the SDSS imaging data to conduct a targeted spectroscopic survey identified seven additional systems
\citep{2009MNRAS.394..367R,2010ApJ...708..456R,Carter:2013fk}.

More recently, a significant number of AM CVn systems has been found from their photometric variability using large-area synoptic surveys. In particular, the Palomar Transient Factory (PTF) has systematically identified seven new AM CVn
systems from their photometric outbursts in a colour-independent manner \citep{2011ApJ...739...68L, Levitan:2013uq,Levitan:1119aq}
as well as over 80 new CVs.
Three AM CVn systems have also been identified in a less systematic fashion from the Catalina Real-Time Transient Survey
\citep[CRTS; ][]{Woudt:2013kx,2014MNRAS.443.3174B}. We note that photometric surveys are only sensitive
to the shorter-period outbursting systems, while spectroscopic surveys are most sensitive to longer-period systems,
which have stronger emission lines.

Despite the significant increase in the known sample, the population density question remains to
be fully answered. \citet{2007MNRAS.382..685R} used the original SDSS sample of AM CVn systems to show that the 
population synthesis estimate by \citet{2001A&A...368..939N} was high by an order of magnitude. 
The re-calibrated population density was used to predict that 40 new systems would be discovered by the follow-up project
\citep{2009MNRAS.394..367R}. Instead, this search yielded only seven new systems, implying that the original population estimates
were a factor of 50 too high \citep{Carter:2013fk}. No explanation for this difference has been given in the literature.

The PTF's search for AM CVn systems has provided a second set of systematically identified systems,
determined without the use of colour-selection, to verify
current population models. However, in order to draw any conclusions on the population
of AM CVn systems from an outburst search, the outburst phenomena itself needs to be better understood.
It is believed that the outburst mechanism in AM CVn systems can be described
by adjustments to the same disc instability model (DIM)
as that used to model the outbursts of CVs (see, e.g., \citealt{2001NewAR..45..449L} for an excellent review). Recent work has, in fact,
shown that the outburst in AM CVn systems can be modeled using the DIM \citep{1997PASJ...49...75T, 2012A&A...544A..13K}, 
although the changes in outburst patterns for AM CVn systems \citep[e.g., CR Boo; ][]{2000MNRAS.315..140K,2001IBVS.5120....1K}
are not yet explained.

Efforts to understand outbursts based on observations have been hampered by the lack of long term light curves
for most AM CVn systems. \citet{2012MNRAS.419.2836R}, hereafter R12, have performed the most substantial work
in this area. They used the Liverpool Telescope to monitor 16 AM CVn systems
for 2.5 years. However, the use of dedicated observations provided only a short baseline,
and even several known outbursting systems were not detected in outburst during their monitoring.
Only a few systems have been monitored for more than a few years \citep[most notably CR Boo; ][]{2013PASP..125..126H},
but the variety of outbursts, as we describe in this paper, requires data for more than one system.

Earlier work on individual systems has provided some information on their outburst recurrence times.
Both \citet{2011ApJ...739...68L}, hereafter L11, and R12 differentiated
between shorter orbital-period systems ($20\,\text{min} < P_{orb} < 27\,\text{min}$) and longer orbital-period
systems ($27\,\text{min} < P_{orb} < 40\,\text{min}$). They noted that the former of these groups
has fairly well established recurrence times of less than a few months while the latter group has either very poorly determined
recurrence times or no determined recurrence time.

Here, we extend the work of R12 by using three separate synoptic surveys to extend our baseline to almost 10\,yrs for many systems.
This allows us, for the first time, to consider the outburst frequency of those systems outbursting only
once every several years. Additionally, since we use non-dedicated observations from large-area surveys,
we are able to analyse recently discovered AM CVn systems by drawing on past data for these systems.
We do note that a significant disadvantage of synoptic surveys is the often erratic coverage and the long cadences.

This paper is organized as follows. We begin by describing the surveys, data processing, and analysis methods in Section \ref{sec:methods}.
We review the known outbursting AM CVn systems in Section \ref{sec:data} and present our composite light curves, 
along with initial analysis of the outbursts.  In Section \ref{sec:discussion}, we discuss AM CVn system evolution, outburst properties,
and make predictions on the observed number of systems in current synoptic surveys.
We summarize our conclusions in Section \ref{sec:summary}.

\section{Observations and Reduction}
\label{sec:methods}

\subsection{Data Sources}

The observations presented in this paper come from three synoptic surveys: the PTF,
the CRTS, and the Lincoln Near Earth Asteroid Research survey (LINEAR).  In the remainder of this section,
we summarize each of these surveys, including an overview of the survey parameters, details of data processing, and
a discussion of the limiting magnitudes presented here for the survey. The limiting magnitudes are
particularly important for this project, as most known outbursting AM CVn systems are extremely faint in quiescence.

\subsubsection{Palomar Transient Factory}
\label{sec:ptf}

The PTF\footnote{\url{http://www.ptf.caltech.edu/}} \citep{2009PASP..121.1395L, 2009PASP..121.1334R} used the Palomar $48"$ Samuel Oschin Schmidt Telescope
to image $7.3\deg^2$ of the sky simultaneously using eleven $2048\times4096$ pixel CCDs.
The typical PTF cadence of 1--5\,d was primarily chosen to discover supernovae. Certain
areas of the sky have been observed with a higher cadence --- from 1 day down to 10 minutes. 
Typically, two individual exposures separated by 30 minutes are taken 
every day to eliminate asteroids and artefacts. The PTF observes in either $R$-band or $g'$-band, with an
$H\alpha$ survey during full moon. The $5\sigma$ limiting magnitude of the survey is
$R\sim20.6$ and $g'\sim21.0$ with saturation around 14$^\text{th}$ magnitude.
The PTF data is the best calibrated and deepest of the large-area synoptic surveys used here. However, it is also the
youngest and has the least amount of data.

The PTF data are processed through the so-called photometric pipeline which uses aperture
photometry and prioritizes photometric accuracy over processing speed \citep{2014PASP..126..674L}. After de-biasing and flat-fielding,
catalogs are generated using Sextractor \citep{1996A&AS..117..393B}. Photometric calibration relative to SDSS fields
observed in the same night provides an absolute calibration accuracy of better than $\simapp2\text{--}3\%$
on photometric nights, but this can be significantly inaccurate on nights with changing
weather conditions \citep{2012PASP..124...62O}. Relative photometric calibration is able to correct for such changes as well
as improve the precision of photometry at the bright end to 6--8\,mmag and at the faint end to $\mathord{\sim}0.2\,$mag.
The basic approach of the algorithm is described in
\citet{2011ApJ...740...65O} and \citet{2011ApJ...739...68L} with PTF-specific details to be published at a future time. 
Although this algorithm is primarily a relative calibration algorithm, it simultaneously uses external calibration references
to provide an absolute calibration. For the PTF data, we use the median value of the absolute-calibrated photometric measurements.

The photometric pipeline produces two limiting magnitude estimates for each exposure as part of the calibration process.
The first estimate defines the limiting magnitude as the magnitude at which 95\% of sources in a deep co-added image
are present in an individual exposure. The second estimate is a theoretical estimate of the maximum magnitude at which a
$5\sigma$ detection is possible. Typically, this $5\sigma$ detection limit is reached $\simapp0.5$\,mag fainter than the 95\%
limiting magnitude estimate, but we have found it to be unreliable in poor weather conditions, in part because it relies
on the zero-points calculated from the comparison to SDSS, which themselves are unreliable in poor weather. Here, we use the
former estimate due to its more consistent performance.

\subsubsection{Catalina Real-Time Transient Survey}

The CRTS\footnote{\url{http://crts.caltech.edu/}} \citep{2009ApJ...696..870D} uses three separate telescopes: the Catalina Sky Survey
0.7\,m Schmidt (CSS), the Mount Lemmon Survey 1.5\,m (MLS), and the Siding Spring Survey 0.5\,m Schmidt (SSS). The fields of
view are, respectively, $8.1\deg^2$, $1.2\deg^2$, and $4.2\deg^2$, with corresponding limiting magnitudes in $V$ of 19.5, 21.5,
and 19.0. The majority of data currently available is from the CSS, and has a typical cadence of one set of 4 exposures per night
per field separated by 10\,min, repeated every 2 weeks.

The CRTS DR2 public release provides both the ability to see all exposures covering a given part of the sky and the ability to download
light curves around a set of coordinates. We began by downloading the list of exposures at each location, as well as the light curve
for the target, from the ``photcat'' catalog. This catalog is the
set of sources identified in deep co-added CRTS images, as part of the CRTS pipeline. 
We retained only those exposures
with $1^{\prime\prime} < FWHM < 4^{\prime\prime}$ and exposure times between 1\,s and 120\,s to eliminate problematic exposures.
We downloaded light
curves of all objects within $\mathord{\sim}0.3\deg^2$ of the centre of the CRTS pointing for these exposures.

Although we would prefer to estimate the limiting magnitude with the same method as that used for PTF exposures,
the lack of publicly available deep co-added images from the CRTS precludes this. We thus estimate the $5\sigma$ limiting magnitude of
each exposure to be the faintest star detected in this set of light curves. We then subtract 0.5\,mag from this limiting magnitude to convert this
into a ``95\% limiting magnitude'', as defined for the PTF (i.e., $m_\mathrm{lim}=m(\mathrm{faintest\;star}) - 0.5$). These estimates are typically consistent with the average limiting magnitudes
of the CRTS \citep{2009ApJ...696..870D}.

A few of the AM CVn systems observed by the CRTS are too faint to be detected in the default ``photcat'' catalog. Detections not associated with this set of sources
are in the ``orphancat'' catalog (A. Drake, priv. comm.). In these
cases, we assumed that any detection in the ``orphancat'' within $3.5^{\prime\prime}$ ($\mathord{\sim}1.5\times$ the pixel scale of the CSS, similar
to criteria used for PTF source association) of the target coordinates was a detection of our target.

\subsubsection{Lincoln Near Earth Asteroid Research survey}

The Lincoln Near Earth Asteroid Research survey\footnote{Public access to
LINEAR data is provided through the SkyDOT web site (\url{https://astroweb.lanl.gov/lineardb/}).}
\citep[LINEAR; ][]{Stokes:2000kx} used two telescopes at the White Sands Missile Range
for a synoptic survey primarily targeted at the discovery of near-Earth objects. 
\citet{Sesar:2011vn} re-calibrated the LINEAR data using the SDSS survey, resulting in $\mathord{\sim}200$ unfiltered observations per object ($\mathord{\sim}600$
observations for objects within $\pm10^\circ$ off the Ecliptic plane) for 25 million objects in the $\sim$9,000 deg$^2$ of sky where
the LINEAR and SDSS surveys overlap (roughly, the SDSS Galactic cap north of Galactic latitude $30^\circ$ and the SDSS Stripe 82 region). Each exposure covered $\mathord{\sim}2\deg^2$ to a $5\sigma$ limiting magnitude of $r'\sim18$, as determined by the
calibration of the unfiltered exposures to the SDSS survey.
The photometric precision of LINEAR
photometry is $\mathord{\sim}0.03$ mag at the bright end ($r' \sim 14$) and $\sim0.2$ mag at $r' = 18$ mag. 

The published LINEAR data set contains information only on source detections, and provides no list of exposures for a particular field.
We thus need to both determine when the target was observed, as well as the limiting magnitudes of those exposures.
To identify exposures on which a particular target was not detected we downloaded light curves for all sources within $20^\prime$ of 
the target. We assumed that a single MJD corresponded to a single exposure and identified those sources for which there were
detections for at least 90\% of the MJDs at which the target was detected. Lastly, we identified all MJDs when this group of
sources was detected and thus found the non-detections of the target
by comparing this list to the list of target detections. 

To estimate limiting magnitudes when the target was not detected, we used a similar
technique as we did with the CRTS data. Since the centre of the frame coordinates is not available, we used only those stars
earlier identified to be near the target. We estimate the 95\% limiting magnitudes to be 0.5\,mag brighter
than the faintest star observed for each exposure.

\subsubsection{Palomar $60^{\prime\prime}$ Data}

Some data for CR Boo were obtained using targeted observations with the Palomar $60^{\prime\prime}$ (P60) telescope.
This data were de-biased, flat-fielded, and astrometrically calibrated with the P60 Automated Pipeline \citep{2006PASP..118.1396C}.
Photometric measurements were made using the \noun{Starlink} package \noun{autophotom} and calibrated using the
relative photometric algorithm described in L11. The absolute scale was tied to the SDSS DR9 catalog
\citep{Ahn:2012zr}.

\subsubsection{Photometric Data Calibration}

Although we use data from three different surveys, we decided to avoid jointly calibrating the light curves.
The primary reason for this decision is that the wide-field nature of the surveys requires a large number of calibration sources.
With the PTF photometric pipeline, we use 350--400 stars to calibrate light curves for each $\mathord{\sim}0.7\deg^2$
section of the sky (that falling on one detector). Given our lack of access to the
raw CRTS and LINEAR data sets, it would be difficult to find this many calibration sources for each target. Although it
is possible to calibrate with fewer stars, the lack of filters for the CRTS and LINEAR surveys makes this calibration more difficult,
since we would need to account for different CCD response curves, the presence of filters, and source colours.
Regardless, our primary interest is in large-scale photometric variability relative to a quiescent magnitude, and even a
systematic offset of several tenths of a magnitude between surveys is acceptable.

\subsection{Outburst Definitions}
\label{sec:outburst_def}
Although outbursts are often easy to identify by eye, a quantitative definition is necessary for a systematic study.
We define an outburst to be $\mathord{\geq}2$ detections that are brighter than the
quiescent magnitude by the greater of 0.5\,mag or $3\sigma\,$mag, where $\sigma$ is the scatter of the light curve
while the system is in quiescence. At least 2 of the detections must be within 15\,d. While the light curve of the system
satisfies both conditions, we consider it to be in outburst. The quiescent magnitude is taken
to be the median of the light curve or, for the faintest systems, from the literature.
Additionally, for PTF, we confirmed all outburst detections by looking
at the individual images. Neither CRTS nor LINEAR images are publicly available at the current time.

We estimate three properties for all outbursting systems presented here: the strength, duration
and recurrence time. 
We define the strength of the outburst to be the difference between the peak luminosity observed and the
quiescent magnitude. This is actually a lower limit on the strength, but without continuous
monitoring it would be difficult to identify the actual peak magnitude. Our estimates for the properties are
consistent with any that exist in the literature.

The outburst duration is even more difficult to determine, due to the infrequent sampling. When available, we used durations from
the literature. When not available, we either estimated or placed an upper limit on the duration using our earlier definition of an outburst.
For systems with multiple, relatively well-sampled outbursts, we used an average of outburst durations.
For systems with only a few observed outbursts and poorly sampled data, we provided an upper limit based on the next detection
not in outburst.

The most difficult to estimate is the recurrence time for those systems for which we observed multiple outbursts. Again, we used
any published estimates if available, except as noted in Section \ref{sec:regular}. For systems with more than five outbursts,
we used the time of the brightest measurement of each outburst, and estimated the recurrence time as their mean. We estimated
the error as the scatter of those measurements around the mean, and assumed that the outbursting behavior remained consistent
throughout any gaps in the data. This implies that the recurrence time is fixed, something known not to be true for at least some
systems, and thus the error will be a combination of inherent variability in the recurrence time and the exact time of observation
at the peak of the outburst. All systems showed a minimum outburst frequency between several outbursts,
and we tested longer gaps
with integer division to check for any observations at the predicted outburst times. PTF1J0719+4858 and CP Eri
showed extra outbursts
that were on time-scales of less than 5\,d and outside of the normal pattern of detections. We assumed these to be normal
outbursts and ignored them for the purposes of estimating the outburst recurrence time.
We generally refrain from using power spectra to
estimate recurrence times due to the irregularity and sparsity of measurements relative to the outburst durations, the
multiple telescopes, and, oftentimes, the lack of detections in quiescence. Shorter period systems do show some signals
corresponding to the observed recurrence times in the power spectra, but these signals are typically weak compared to the noise.

For those systems showing fewer than five outbursts, we estimated the recurrence time as the average time
between outbursts. We assigned errors based on a propagation from the uncertainty of duration in the few outbursts observed
(i.e., the time from previous observation to observation in outburst), but we emphasize that the few outbursts seen make any
error estimation difficult. We tested whether the recurrence time could be our estimate divided by an integral value by looking
for observations at the predicted times (a simplistic use of the standard $O-C$ technique). We remark on any adjustments as
part of our individual system descriptions in Section \ref{sec:regular}.

\section{AM CVn Systems and Observational Data}

\label{sec:data}

We present the known AM CVn systems in Table \ref{tbl:amcvns}, along with some information on data sources and the
presence of outbursts. In this paper, we present only light curves showing significant variability. Combined light curves for all
systems, including those which show no variability, are available from the PTF website\footnote{\url{http://ptf.caltech.edu/}}.
Here, we differentiate between three behavioral classes: those systems showing repeated outbursts, those with a single observed outburst,
and those with irregular photometric behavior.

\begin{table*}
\centering
\begin{minipage}{\textwidth}
\caption{Known AM CVn Systems.}
\label{tbl:amcvns}
\begin{tabular}{lccccccccc}
\hline
System$^{a}$ & Outbursting & Period & Quiescence & PTF$^b$ & CSS$^b$ & MLS/SSS$^{b,c}$ &  LINEAR$^b$ & References \\
&&(min)&(g')&&&&&\\
\hline

HM Cnc & N & 5.36 & 20.7 & 58/59 & \nodata & \nodata & \nodata & 1 \\
V407 Vul & N & 9.48 & 19.7 & \nodata & \nodata & \nodata & \nodata & 2 \\
ES Ceti & N & 10.3 & 17.1 & \nodata & 164/235 & \nodata & \nodata & 3 \\
KIC 004547333 & N & 15.9 & 16.1 & 117/118 & 31/36 & \nodata & \nodata & 4 \\
AM CVn & N & 17.1 & 14.2 & 103/104 & \nodata & \nodata & 293/293 & 5 \\
HP Lib & N & 18.4 & 13.5 & \nodata & 131/134 & S: 130/130 & \nodata & 6 \\
PTF1\,J191905.19+481506.2 & Y & 22.5 & 21.5 & 22/110 & \nodata & \nodata & \nodata & 7 \\
CR Boo & Y & 24.5 & 17.4 & 31/31 & 286/286 & \nodata & 266/271 & 8, 9 \\
KL Dra & Y & 25.0 & 19.1 & \nodata & \nodata & \nodata & \nodata & 10 \\
V803 Cen & Y & 26.6 & 16.9 & \nodata & \nodata & S: 231/231 & \nodata & 6, 11, 12 \\
PTF1\,J071912.13+485834.0 & Y & 26.8 & 19.4 & 250/262 & 281/292 & \nodata & \nodata & 13 \\
SDSS\,J092638.71+362402.4 & Y & 28.3 & 19.0 & 8/8 & 254/295 & \nodata & 77/714 & 14, 15 \\
CP Eri & Y & 28.7 & 20.3 & 198/300 & 160/228 & S: 35/45 & \nodata & 16 \\
PTF1\,J094329.59+102957.6 & Y & 30.4 & 20.7 & 71/217 & 50/296 & M: 51/53 & 16/1163 & 17 \\
V406 Hya & Y & 33.8 & 20.5 & \nodata & 83/262 & \nodata & \nodata & 18 \\
PTF1\,J043517.73+002940.7 & Y & 34.3 & 22.3 & 2/213 & 7/319 & \nodata & \nodata & 17 \\
SDSS\,J173047.59+554518.5 & N & 35.2 & 20.1 & \nodata & 69/119 & \nodata & 0/535 & 19, 20 \\
2QZ\,J142701.6--012310 & Y & 36.6 & 20.3 & \nodata & 62/298 & \nodata & 19/493 & 21 \\
SDSS\,J124058.03--015919.2 & Y & 37.4 & 19.7 & \nodata & 224/302 & M: 86/86 & 39/529 & 22 \\
SDSS\,J012940.05+384210.4 & Y & 37.6 & 19.8 & \nodata & 74/260 & \nodata & \nodata & 14, 23, 24 \\
SDSS\,J172102.48+273301.2 & Y & 38.1 & 20.1 & 208/298 & 31/382 & \nodata & 0/409 & 25, 26 \\
SDSS\,J152509.57+360054.5 & N & 44.3 & 19.8 & 80/100 & 181/254 & \nodata & 60/231 & 24, 25 \\
SDSS\,J080449.49+161624.8 & \nodata$^d$ & 44.5 & 18.2 & 110/112 & 336/358 & \nodata & \nodata & 27 \\
SDSS\,J141118.31+481257.6 & N & 46.0 & 19.4 & 102/111 & 84/121 & \nodata & 0/237 & 14 \\
GP Com & N & 46.5 & 15.9 & 11/12 & 315/315 & \nodata & 207/450 & 28 \\
CRTS\,J045020.8--093113 & Y & 47.3 & 20.5 & 31/66 & 21/240 & \nodata & \nodata & 29 \\
SDSS\,J090221.35+381941.9 & Y$^e$ & 48.3 & 20.2 & \nodata & 47/341 & \nodata & 0/337 & 25, 30 \\
SDSS\,J120841.96+355025.2 & N & 52.6 & 18.8 & 97/101 & 283/288 & \nodata & 101/290 & 24, 31 \\
SDSS\,J164228.06+193410.0 & N & 54.2 & 20.3 & \nodata & 1/369 & \nodata & 0/430 & 24, 25 \\
SDSS\,J155252.48+320150.9 & N & 56.3 & 20.2 & 125/242 & 47/297 & \nodata & 0/230 & 32 \\
SDSS\,J113732.32+405458.3 & N & 59.6 & 19.0 & 72/77 & 300/309 & \nodata & 0/539 & 33 \\
\smallskip
V396 Hya & N & 65.1 & 17.3 & 54/56 & 46/48 & S: 235/236 & \nodata & 34 \\
SDSS\,J150551.58+065948.7 & N & \nodata & 19.1 & 143/149 & 337/347 & \nodata & 106/606 & 33 \\
CRTS\,J084413.6--012807 & Y & \nodata & 20.3 & \nodata & 22/324 & \nodata & \nodata & 35 \\
SDSS\,J104325.08+563258.1 & Y & \nodata & 20.3 & 14/16 & 22/120 & \nodata & 34/216 & 19 \\
PTF1\,J221910.09+313523.1 & Y & \nodata & 20.6 & 49/72 & 53/111 & \nodata & \nodata & 17 \\
CRTS\,J074419.7+325448 & Y & \nodata & 21.1 & \nodata & 103/460 & M: 32/49 & \nodata & 35 \\
PTF1\,J085724.27+072946.7 & Y & \nodata & 21.8 & 15/126 & 50/349 & \nodata & 0/791 & 17 \\
PTF1\,J163239.39+351107.3 & Y & \nodata & 23.0 & 61/173 & 36/324 & \nodata & 0/564 & 17 \\
PTF1\,J152310.71+184558.2 & Y & \nodata & 23.5 & 10/28 & 2/325 & \nodata & 0/203 & 17 \\
SDSS\,J204739.40+000840.1 & Y & \nodata & 24.0 & \nodata & 0/67 & \nodata & 0/591 & 31 \\
\hline
\end{tabular}
\smallskip\\
Systems are sorted by orbital period. System with no orbital period in the literature are at the bottom and sorted by quiescence magnitude.\\
$^a$ Names given here are either the IAU variable star name or the full name given in the discovery paper. Throughout this
paper, we use a shortened version of the latter.\\
$^b$ Survey columns are of the form `\# of detections / \# of observations'. \\
$^c$ Since no system has observations from both MLS and SSS, we use one column for both surveys and indicate the appropriate survey.\\
$^d$ SDSSJ0804+1616 has non-outburst variability. See Section \ref{sec:other}.\\
$^e$ SDSSJ0902+3819 was recently reported to outburst \citep{2014arXiv1407.4196K}. Our data here does not include this outburst.\\
\smallskip\\
References: (1) \citet{2010ApJ...711L.138R}; (2) \citet{Steeghs:2006ve}; (3) \citet{Espaillat:2005qf}; (4) \citet{Fontaine:2011ly}; (5) \citet{2006MNRAS.371.1231R}; (6) \citet{2007MNRAS.379..176R}; (7) \citet{Levitan:1119aq}; (8) \citet{1997PASP..109.1100P}; (9) \citet{2000MNRAS.315..140K}; (10) \citet{Ramsay:2010uq}; (11) \citet{2000PASP..112..625P}; (12) \citet{Kato:2004vn}; (13) \citet{2011ApJ...739...68L}; (14) \citet{2005AJ....130.2230A}; (15) \citet{2011MNRAS.410.1113C}; (16) \citet{2001ApJ...558L.123G}; (17) \citet{Levitan:2013uq}; (18) \citet{2006MNRAS.365.1109R}; (19) \citet{Carter:2013fk}; (20) \citet{2014MNRAS.437.2894C}; (21) \citet{2005IAUC.8531....3W}; (22) \citet{2005MNRAS.361..487R}; (23) \citet{2011arXiv1104.0107S}; (24) \citet{2013MNRAS.432.2048K}; (25) \citet{2010ApJ...708..456R}; (26) Augusteijn, T, private communication; (27) \citet{2009MNRAS.394..367R}; (28) \citet{1981ApJ...244..269N}; (29) \citet{Woudt:2013kx}; (30) \citet{2014arXiv1407.4196K}; (31) \citet{2008AJ....135.2108A}; (32) \citet{Roelofs:2007dq}; (33) \citet{2014MNRAS.439.2848C}; (34) \citet{2001ApJ...552..679R}; (35) \citet{2014MNRAS.443.3174B}

\end{minipage}
\end{table*}

\subsection{Regularly Outbursting Systems}
\label{sec:regular}

\begin{figure*}
\includegraphics{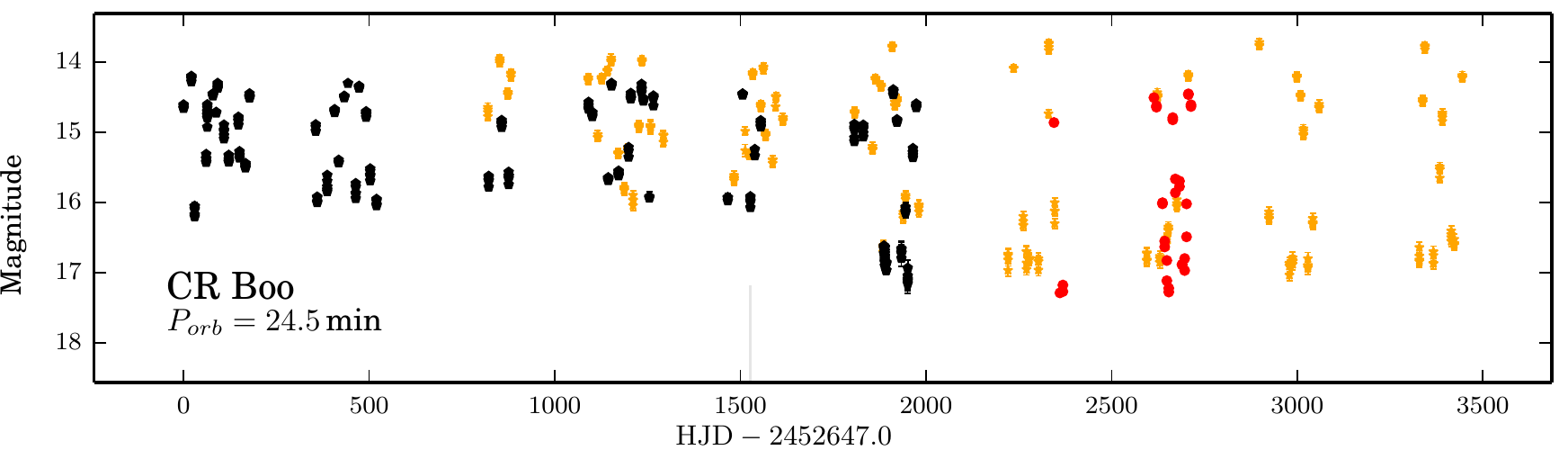}\\
\includegraphics{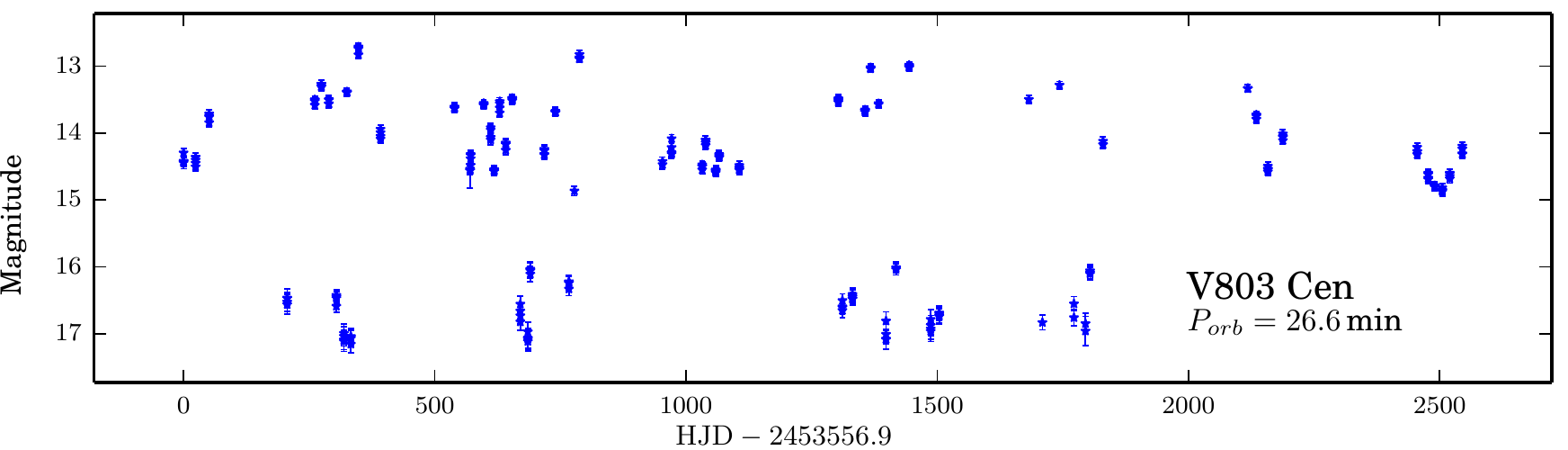}\\
\includegraphics{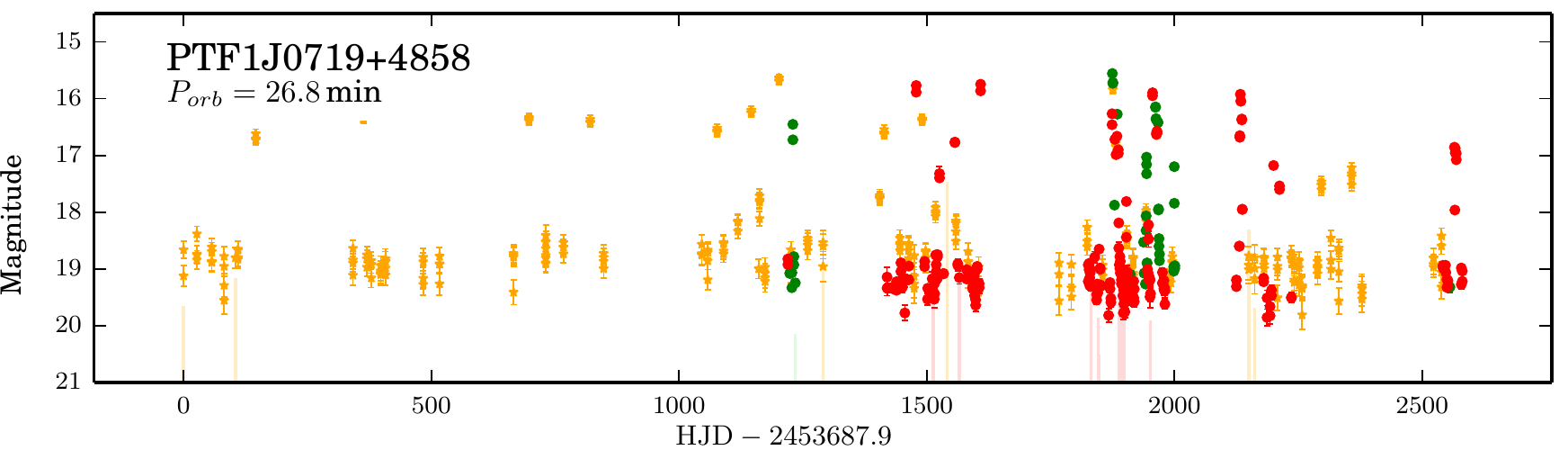}\\
\includegraphics{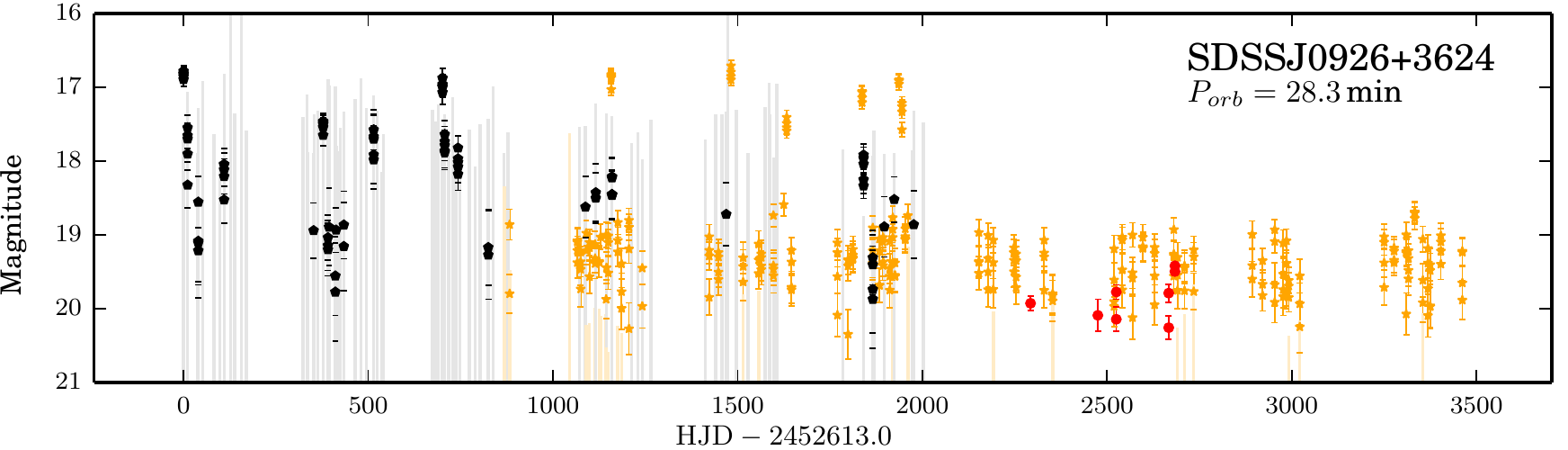}\\
\caption{Light curves of the four shortest-period regularly outbursting AM CVn systems presented here. All show regular
changes from quiescence to outburst (Section \ref{sec:regular}). In particular, we point out the significant change in the behavior of CR Boo (Section \ref{sec:crboo})
and of SDSSJ0926+3624 (Section \ref{sec:sdss0926}).\newline
\textbf{Legend:} black = LINEAR; yellow = CSS; blue = SSS; red = PTF $R$; green = PTF $g'$. The tops of the vertical lines (colour-coded to match the survey) are limiting magnitudes for non-detections.
\label{fig:lcs:regular1}}
\end{figure*}

\begin{figure*}
\includegraphics{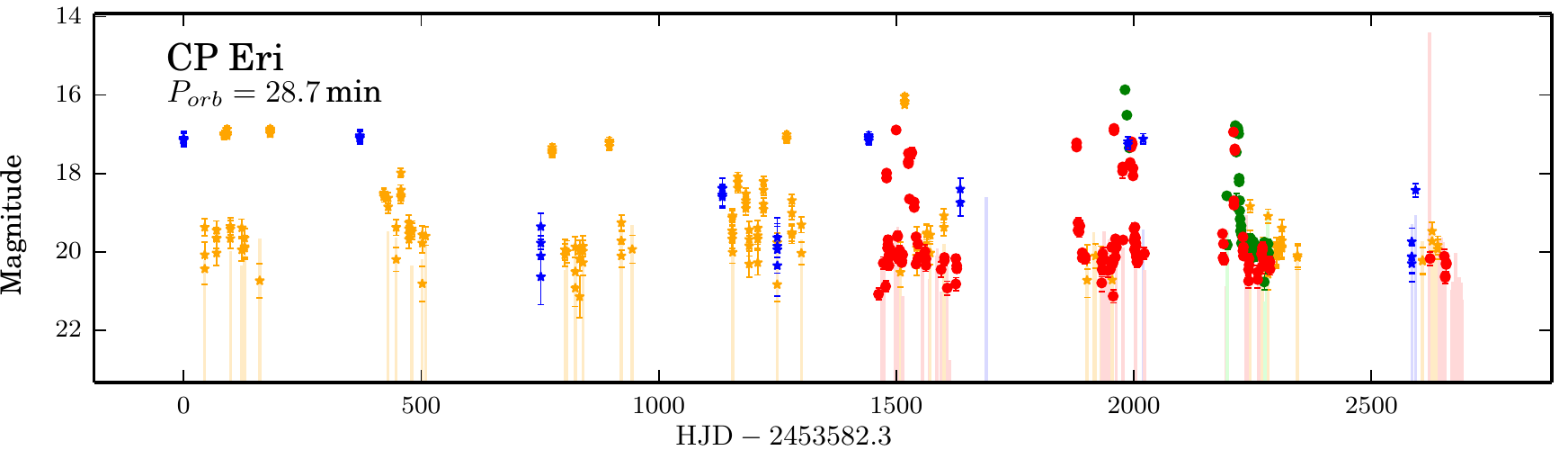}\\
\includegraphics{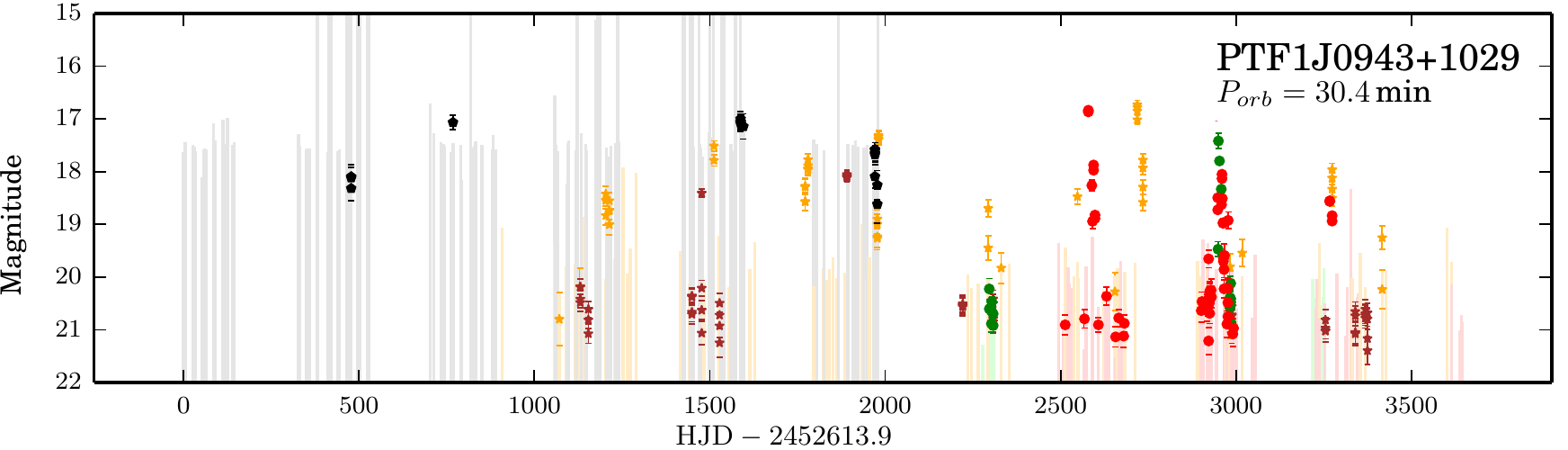}\\
\includegraphics{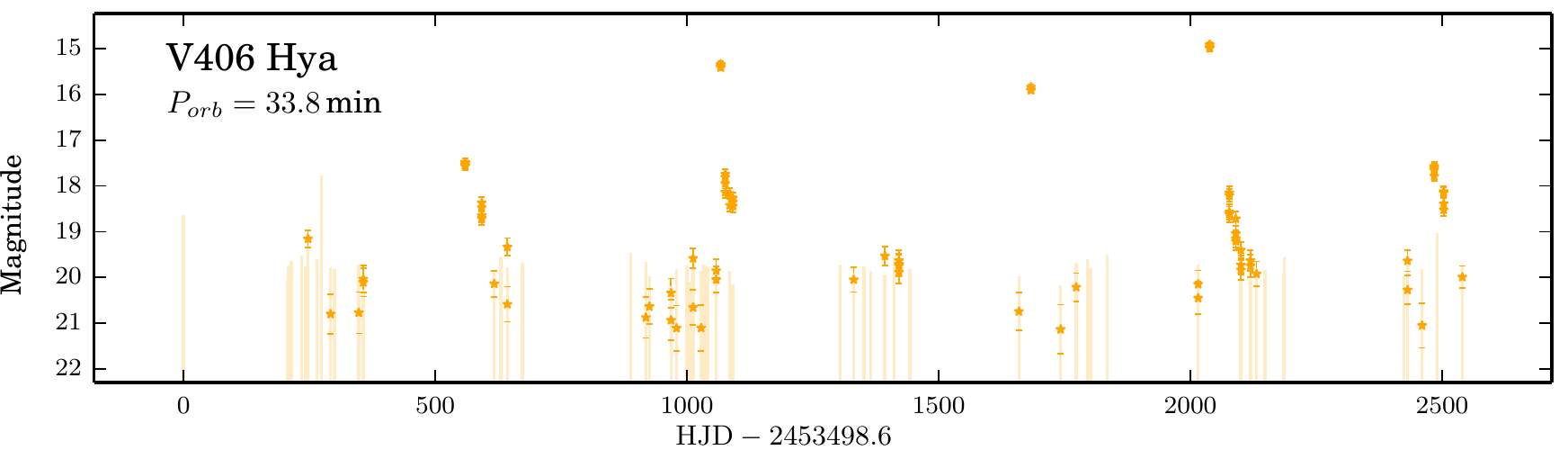}\\
\caption{Light curves of three regularly outbursting AM CVn systems in order of $P_{orb}$, which all show regular
changes from quiescence to outburst (Section \ref{sec:regular}).  In contrast with the light curves in Figure \ref{fig:lcs:regular1},
all systems in this figure spend the majority of their time in quiescence with only occasional outbursts. This is particularly true
for V406 Hya. \newline
\textbf{Legend:} black = LINEAR; yellow = CSS; blue = SSS; maroon = MLS; red = PTF $R$; green = PTF $g'$. The tops of the vertical lines (colour-coded to match the survey) are limiting magnitudes for non-detections.
\label{fig:lcs:regular2}}
\end{figure*}

\begin{figure*}
\includegraphics{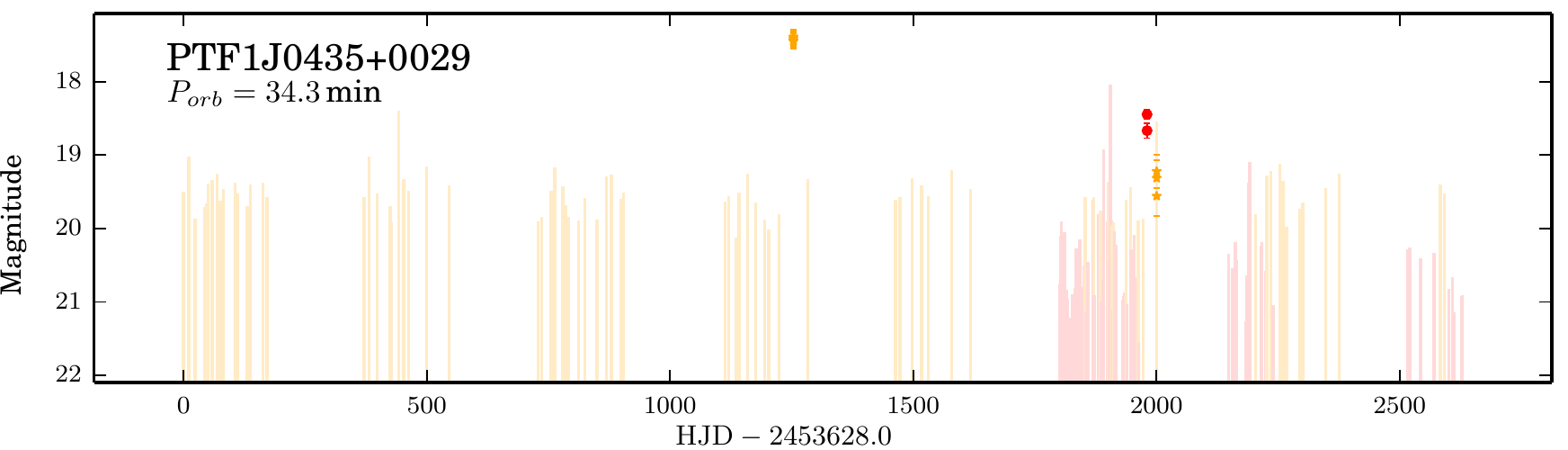}\\
\includegraphics{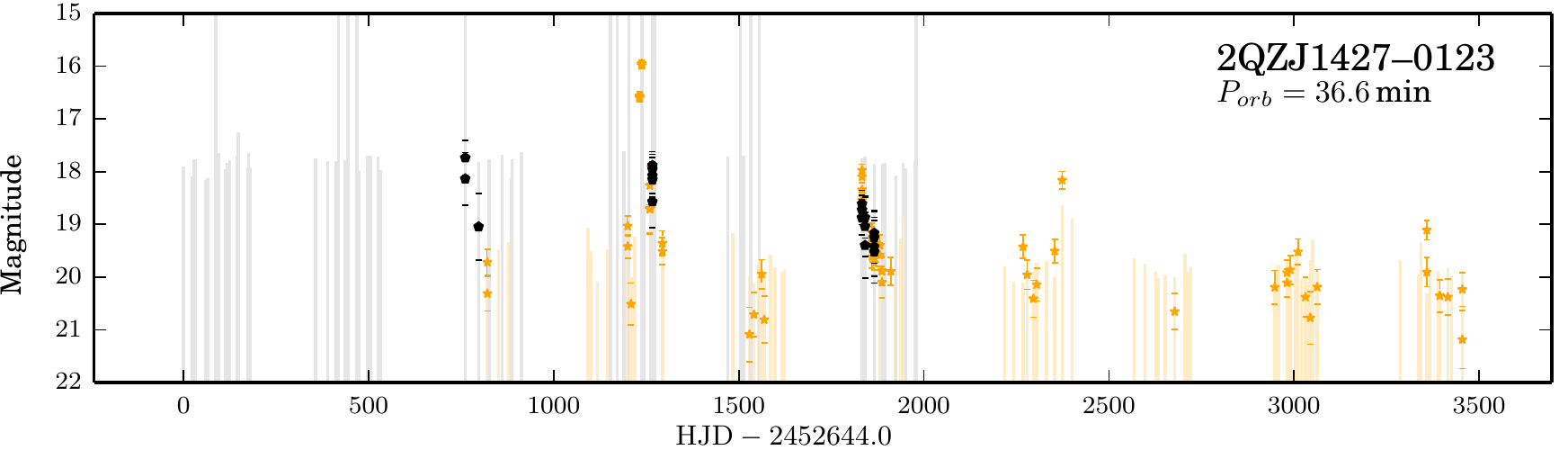}\\
\caption{Light curves of the two longest-period, known, regularly outbursting AM CVn systems. Both systems show only a few outbursts with
recurrence times of $\mathord{\geq}1\,$yr\newline
\textbf{Legend:} black = LINEAR; yellow = CSS; red = PTF $R$. The tops of the vertical lines (colour-coded to match the survey) are limiting magnitudes for non-detections.
\label{fig:lcs:regular3}}
\end{figure*}

\begin{figure*}
\includegraphics{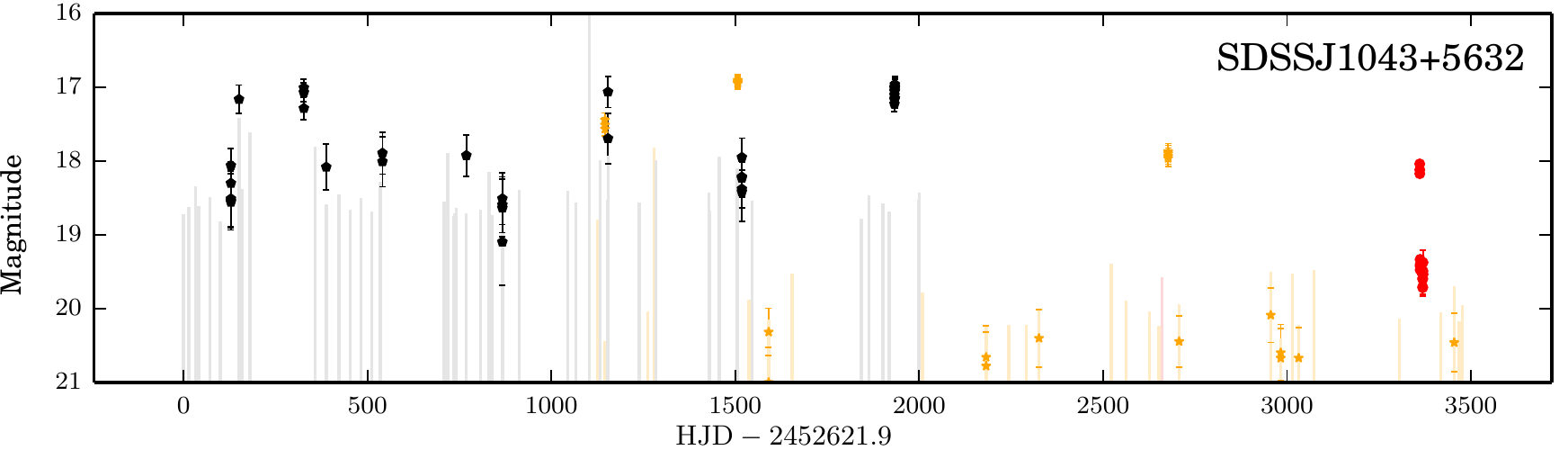}\\
\includegraphics{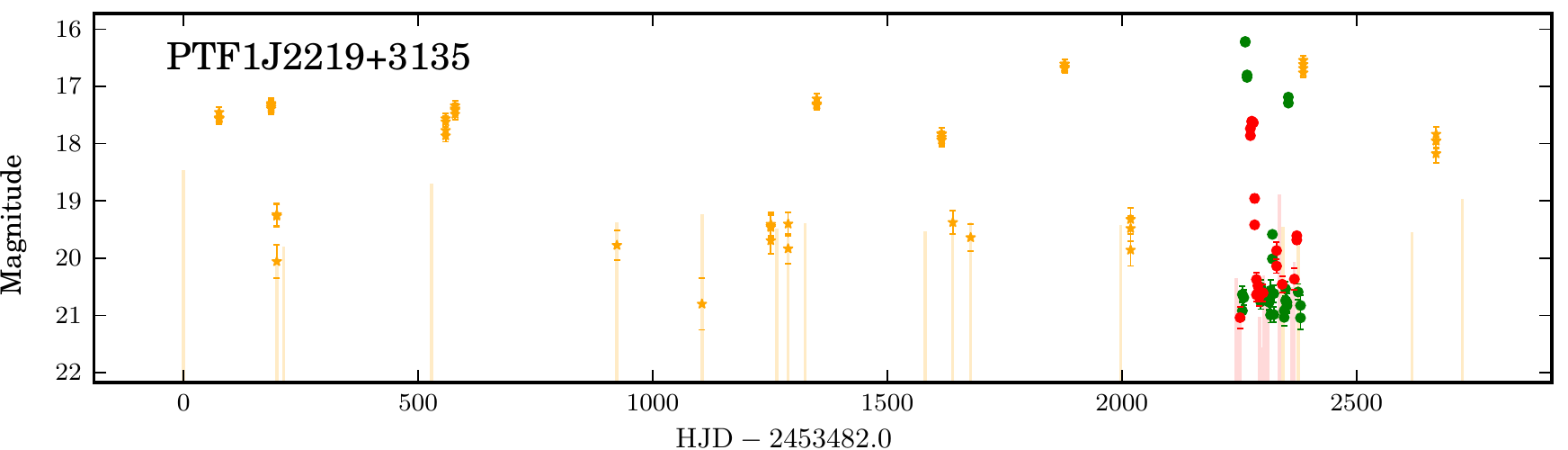}\\
\includegraphics{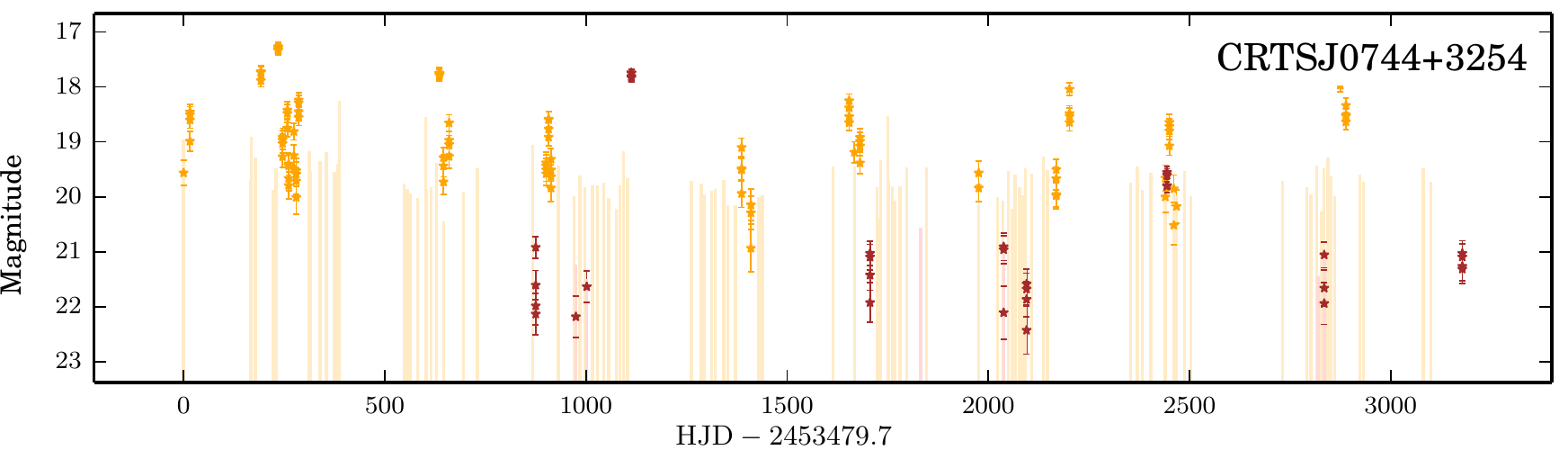}\\
\includegraphics{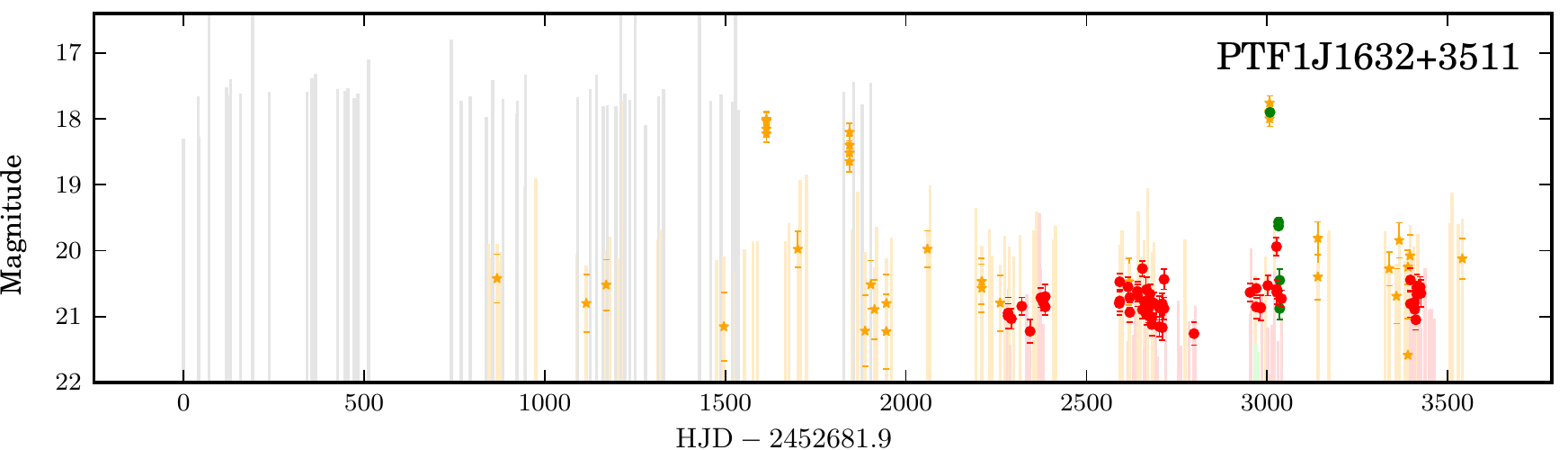}
\caption{Light curves of four regularly outbursting AM CVn systems with unknown orbital periods. We use their outburst recurrence times
to estimate orbital periods in Section \ref{sec:prediction}.\newline
\textbf{Legend:} black = LINEAR; yellow = CSS; red = PTF $R$; green = PTF $g'$. The tops of the vertical lines (colour-coded to match the survey) are limiting magnitudes for non-detections.
\label{fig:lcs:regular4}}
\end{figure*}

\begin{figure*}
\includegraphics{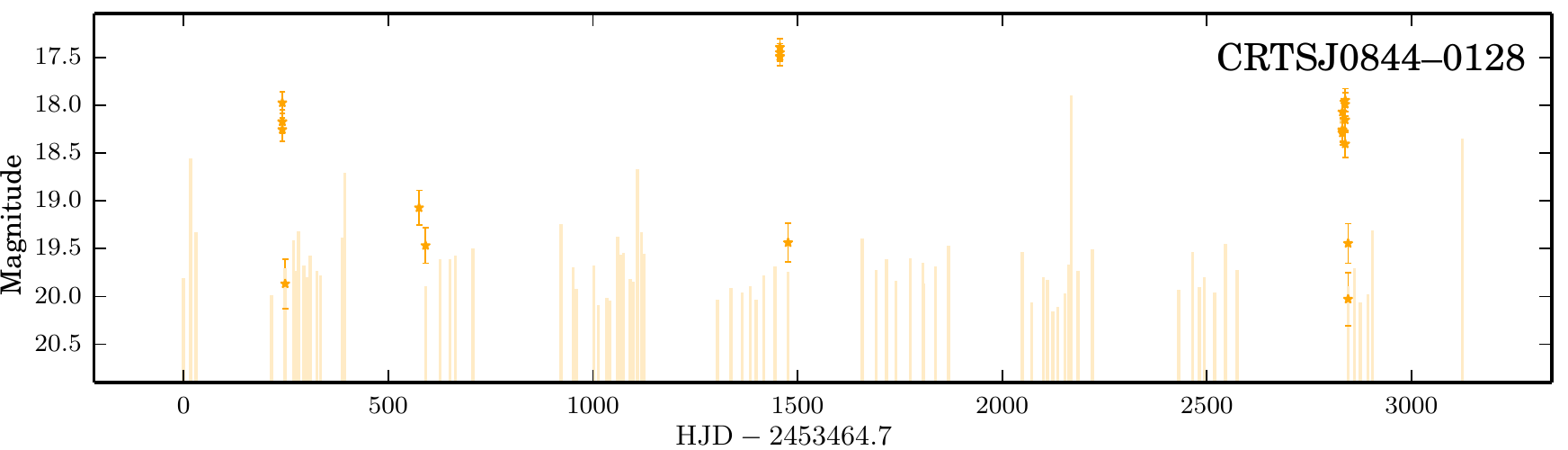}\\
\includegraphics{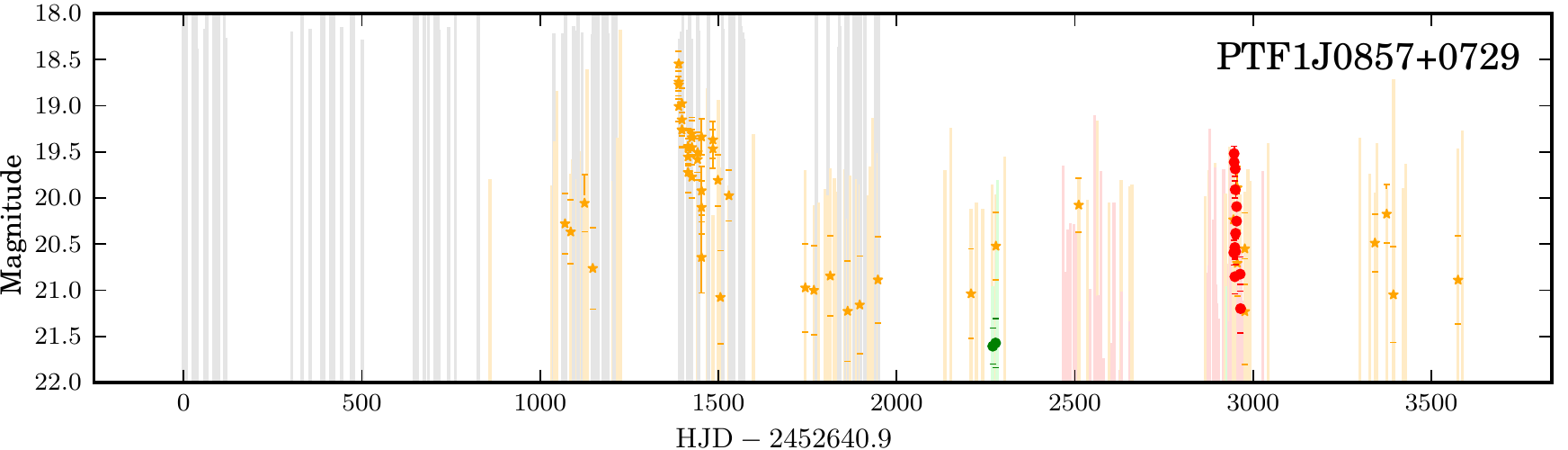}
\caption{Light curves of two regularly outbursting AM CVn systems with unknown orbital periods and extremely long outburst recurrence times. These are discussed in Section \ref{sec:longrecuroutbursts}.\newline
\textbf{Legend:} black = LINEAR; yellow = CSS; red = PTF $R$; green = PTF $g'$. The tops of the vertical lines (colour-coded to match the survey) are limiting magnitudes for non-detections.
\label{fig:lcs:regular5}}
\end{figure*}

In Figures \ref{fig:lcs:regular1}, \ref{fig:lcs:regular2}, \ref{fig:lcs:regular3}, \ref{fig:lcs:regular4}, and \ref{fig:lcs:regular5} we present outburst light curves of 15 systems with multiple observed outbursts.
Two systems known to outburst frequently, PTF1J1919+4815 and KL Dra, are not presented here due to lack of data in the currently
discussed surveys, but we refer the reader to \citet{Levitan:1119aq} and \citet{Ramsay:2010uq}, respectively, for detailed analysis of 
their light curves.  We used the outburst criteria detailed in Section \ref{sec:outburst_def} to identify outbursts in a quantitative fashion, and provide summary data of the outburst characteristics
in Table \ref{tbl:recurringoutbursts}. We provide more in-depth discussion on selected systems below. All outburst times are relative to the
start of the light curve, which is indicated in the respective figure.

\subsubsection{CR Boo}
\label{sec:crboo}

CR Boo was found to have a 46.3\,d outburst recurrence time by \citet{2000MNRAS.315..140K}, hereafter K00.
However, \citet{2001IBVS.5120....1K}, hereafter K01,
reported that this was not constant and that CR Boo had switched to a 14.7\,d recurrence time in 2001. More recent
work by \citet{2013PASP..125..126H}, hereafter H13, presents twenty years of CR Boo photometry and also shows
significant changes in its photometric behavior. The more than
nine years of regular monitoring presented here provides a complementary view of CR Boo's behavior, particularly
in the time period since 2004 when H13's sampling is much more irregular.

The most surprising feature of the long-term light curve presented is a clear distinction in behavior between the first $\mathord{\sim}4.5\,$yrs
and the remaining data (Figure \ref{fig:lcs:regular1}). We will refer to these separate parts of the light curve as the ``active'' and
``inactive'' states. In the active state ($2452647<HJD<2454337$), CR Boo was only observed between $14<V<16$.
In contrast, during the inactive state
($2454337<HJD<2456147$), CR Boo was observed near
its quiescent state ($V<16$) $\mathord{\sim}50\%$ of the time. The abrupt change in behavior is present in both the LINEAR and CSS data.

\begin{figure}
\includegraphics{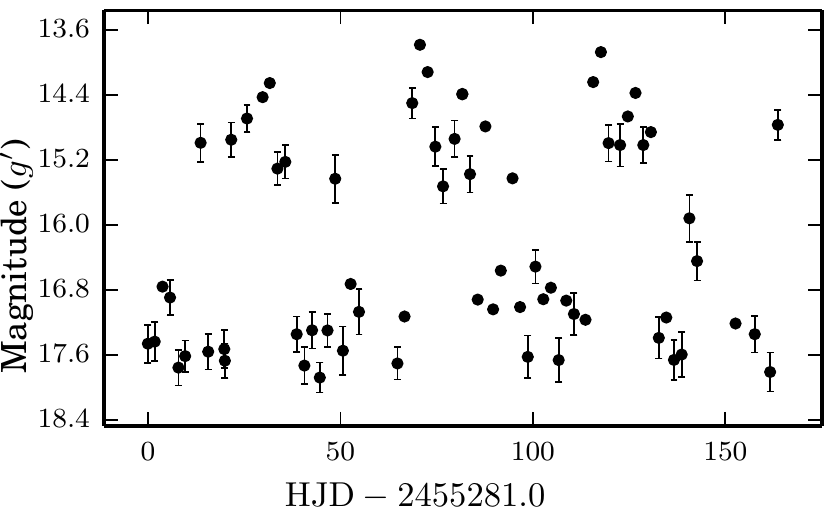}\\
\includegraphics{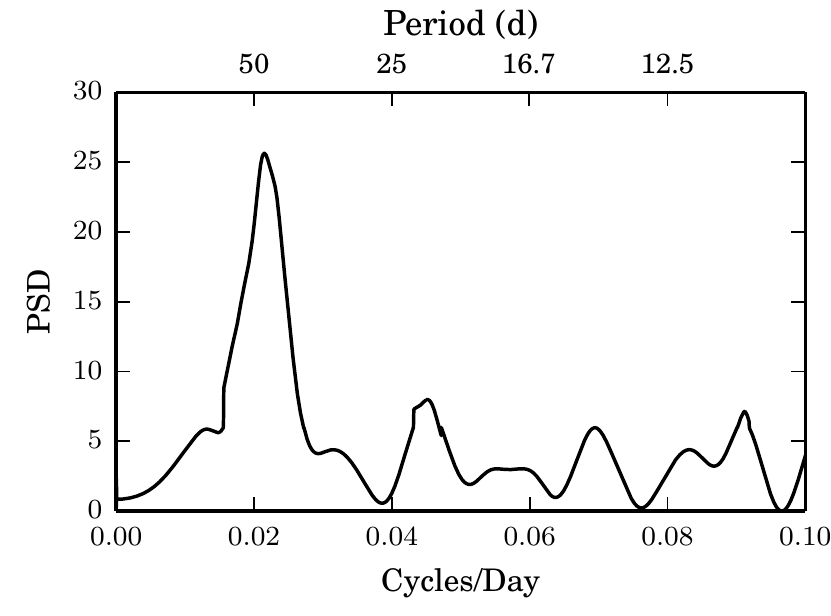}\\
\includegraphics{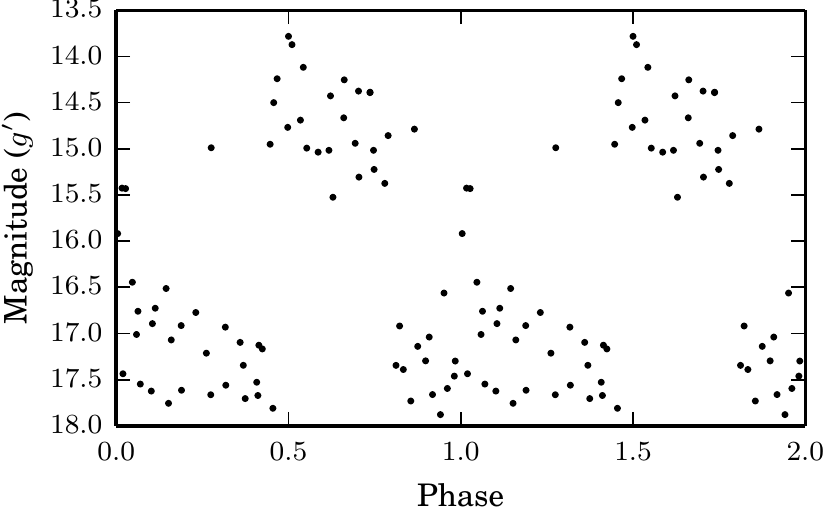}\\
\caption{\textbf{Top:} The un-folded light curve of CR Boo taken by the P60. Three outbursts are clearly visible.
We use this much higher-cadence and more regular light curve to establish that a period of $\mathord{\sim}50\,$d
is real.\newline
\textbf{Middle:} A periodogram of the CR Boo P60 data, showing a peak at 46.5\,d.\newline
\textbf{Bottom:} The CR Boo P60 data light curve folded at the peak period of 46.5\,d, with the peak of the outburst set to
a phase of 0.5. The outburst and quiescent
portions of the light curve are clearly separated.}
\label{fig:crboop60}
\end{figure}

Although an obvious step is to search for periodicity in the data, the infrequent and uneven sampling of the CSS and LINEAR
surveys prevents a comprehensive analysis. Without compelling evidence, even a peak with significant power in a periodogram
may be false. Instead, we consider the recurrence time during CR Boo's inactive state by using a set of observations from the P60 that were taken
over $\mathord{\sim}160\,$d and with a nominal cadence of 3\,d. This provides a much better data set for period analysis.
The peak of the periodogram for the P60 observations is at $46.5\,$d. This estimate is consistent with the outburst
recurrence time found by K00. We present these observations, a periodogram generated from them, and a folded light curve in Figure \ref{fig:crboop60}.

We estimate an error of $10.5\,$d for this period by a bootstrap process \citep{1982jbor.book.....E}. To calculate the error, we drew, at random,  68 observations
from the total set of 68 observations, allowing for repetition. This randomizes both the number of observations and which observations
are used. We then calculated a Lomb-Scargle periodogram \citep{1982ApJ...263..835S} for the randomly drawn data, and recorded the peak. We repeated this process
500 times, and used the standard deviation as the error estimate.

\begin{figure}
\includegraphics{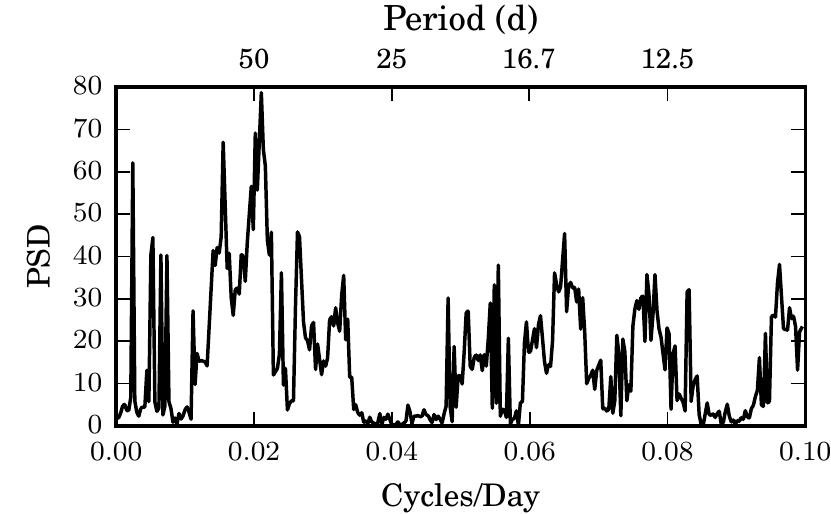}\\
\includegraphics{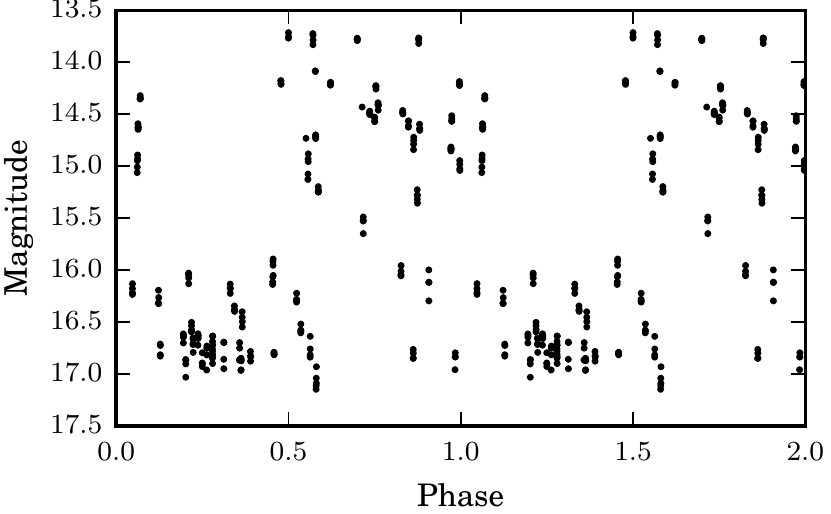}
\caption{\textbf{Top:} A periodogram of the CSS and LINEAR data of CR Boo. The strongest peak is at 47.6\,d, with an associated error
of 4.8\,d. Only with the proof from the P60 data in Figure \ref{fig:crboop60} do we believe that this is a real period.\newline
\textbf{Bottom:} A folded light curve of the CSS and LINEAR data of CR Boo's while it is in its inactive state. The data are folded
at the above period of 47.6\,d, and shows a clear outburst and quiescent states. The recurrence time is consistent over 5\,yrs.\label{fig:crboo_inactive}}
\end{figure}

We now use the much more extensive data for CR Boo from the LINEAR and the CSS surveys, and again compute a periodogram.
Here we have a peak at $47.6\,\text{d}\pm4.8\,$d. We present a periodogram and folded light curve in Figure \ref{fig:crboo_inactive}.

The outburst recurrence time is statistically consistent between the P60 observations, the LINEAR and CSS observations, and the earlier work
by K00 and H13. In particular, H13 found a dominant spacing between outbursts of 46\,d over 20\,yrs.
It is thus likely that the dominant outburst recurrence time 
is the same between active and inactive states and is around 46\,d. For the analysis in this paper, we use the value we derived
from the LINEAR and CRTS data, as it is derived from 5\,yrs of data.

Our data are in agreement with those of H13, specifically regarding the changing state of CR Boo. However, H13 shows even more
variability in the long-term light curve, particularly during the time period that is not covered by the data presented here (1990--2000).
We believe that CR Boo's inactive state between 2005 and 2010 has been remarkably stable, particularly given the relatively
clean outburst light curves presented here. It is obvious that the system often experiences rapid changes in its behavior.

\subsubsection{V803 Cen}

V803 Cen was found by \citet{Kato:2004vn} to have a 77\,d outburst recurrence time with very similar characteristics to the active state of CR Boo described in Section \ref{sec:crboo}. In contrast to CR Boo, the light curve presented here (Figure \ref{fig:lcs:regular1}) shows no
significant changes in the amplitude of photometric variability over almost 7\,yrs. We see no coherent light curve
when folded at the recurrence time given by \citet{Kato:2004vn}. No significant period in a periodogram calculated from the SSS data results in a coherent light curve either, which is consistent with the data of CR Boo in its active state. We thus use the period found by \citet{Kato:2004vn} for our analysis in this paper and assume a 10\% error, consistent with the variability in the outburst recurrence times of
CR Boo, KL Dra, and PTF1J0719+4858 (see Table \ref{tbl:amcvns} for references). It is possible that this lack of periodicity is
due to changing outburst recurrence times, as seen for CR Boo during its active state (Section \ref{sec:crboo}).

\subsubsection{SDSSJ0926+3624}
\label{sec:sdss0926}

SDSSJ0926+3624 is perhaps the best understood AM CVn system, given its deep eclipses. \citet{2011MNRAS.410.1113C}
reported on two outbursts and showed the CSS light curve. The light curve we present here has both additional historical data
from the LINEAR survey, as well as newer data from the CSS. Similarly to CR Boo, SDSSJ0926+3624 shows a dramatic change in behavior
roughly half way through the light curve (Figure \ref{fig:lcs:regular1}).
The earlier part of the light curve ($HJD \lesssim 2461620$) shows repeated outbursts, with a recurrence time
of 140--180\,d. 

The latter part of the light curve ($HJD \gtrsim 2461620$), however, does not show any outbursts. Given that the cadence of CSS did not change,
this is surprising, and is likely an indicator of a real change in the system. We do know that at least one outburst was
missed in the CSS coverage --- that reported in \citet{2011MNRAS.410.1113C} to have occurred in March 2009. Although
it is possible that others were missed as well, we estimate only a $\mathord{\sim}4\%$ probability of a missed outburst, based on
times between CSS observations, the expectation of an outburst at least every 180 days with a duration of at least 20 days,
and not accounting for any particular pattern of outburst relative to the previous outburst. This implies that the outburst
behavior of SDSSJ0926+3624  likely changed, whether to less frequent outbursts or ones that return to quiescence faster.

\subsubsection{CP Eri}
\label{sec:cperi}

Previous studies of AM CVn systems have identified only a few outbursting  systems that show both
super outbursts and normal outbursts. These systems (PTF1J1919+4815, CR Boo, and PTF1J0719+4858) have 
some of the shortest known orbital periods of the outbursting systems.
The normal outbursts are typically 1--2 days in length and appear to have a similar or slightly lower strength as
super-outbursts (e.g., K00, L11). The data presented here show that CP Eri (Figure \ref{fig:lcs:regular2}), a slightly
longer-period system with $P_{orb}=28.7\,$min, also appears
to show normal outbursts. Three increases in brightness of at least two magnitudes between super outbursts
are constrained to last fewer than five days --- consistent with what would be expected for a normal outburst.
This likely indicates that other longer-period AM CVn systems also show normal outbursts in addition to super outbursts.

\subsubsection{PTF1J0435+0029}

In seven yrs of coverage with the CSS and the PTF, PTF1J0435+0029 was observed in outburst twice (Figure \ref{fig:lcs:regular3}). Given the faint nature
of the system, only an observation at the very beginning of the outburst would be above the limiting magnitude of both surveys, and thus the lack of
additional outbursts is not surprising. The two observed outbursts were $\mathord{\sim}730\,$d apart ($t=1250\pm30\,$d and $1980^{+50}_{-8}\,$d),
but the time half way between had no observations, and hence
both 365\,d and 730\,d recurrence times are consistent with the data. Here, we use the former, as the latter would be a significant outlier from the
remainder of the AM CVn systems (see Table \ref{tbl:recurringoutbursts}). Only further observations can remove this ambiguity.

\subsubsection{2QZ\,J1427-01}

We find three outbursts for 2QZ J1427--01 (Figure \ref{fig:lcs:regular3}), with peak magnitudes at $t=760^{+40}_{^-50}\,$d,
$1240^{+30}_{-20}\,$d, and $1830^{+10}_{-30}\,$d. We constrain the duration of the outbursts to $\mathord{<}50\,$d, based on the
second outburst. We provide estimates for the remaining two outbursts using this outburst duration to obtain a lower bound
on their times of peak luminosity, since both outbursts occurred before the start of an observing season.
The mean difference between these peaks is $540\pm65$\,d, with the error
derived based on the errors of each outburst peak. We note that this is roughly consistent with the 10--20\% change in
outburst recurrence time observed in shorter period systems.

These outbursts occur over a period of $\mathord{\sim}1000$\,d, while we have data over a timespan of $\mathord{>}3500$\,d.
We thus expect additional outbursts at $t\approx$ 210\,d, 2370\,d, 2910\,d, and 3450\,d. The first falls between observing
seasons, while the third and fourth are just before and after an observing season, respectively. Given the associated error,
it is highly likely that no outburst would have been seen. There are observations at $t=2354\,$d, 2374\,d, and 2401\,d, roughly
coincident with when we would expect an outburst. One of the exposures on $t=2374\,$d does show a detection consistent
with an outburst, while the remaining three exposures do not indicate outbursts. This may indicate that the system was at
the end of an outburst. We note that the data obtained by R12 does not provide coverage of these predicted outburst times.

We also consider whether the outburst recurrence time may be shorter. A recurrence time of one-half the proposed value
would require outbursts at $t=1560\,$d and 2640\,d, both of which are in the middle of observing seasons. Likewise, one-third
of the proposed value also shows coverage during times of expected outbursts.
We thus conclude that 2QZ\,J1427-01 has an outburst recurrence time of $540\pm65\,$d.

\subsubsection{CRTSJ0804--0128 and PTF1J0857+0729}
\label{sec:longrecuroutbursts}

Two systems, CRTSJ0804--0128 and PTF1J0857+0729, have only a few recorded outbursts but with other observations
almost at the level of an outburst. The outburst recurrence time for the former is approximately 1300\,d while for the latter it
is approximately 1550\,d. Such recurrence times are not similar to the other systems presented here. Given the lack
of measured orbital periods for either system, we do not know if these long outburst recurrence times indicate much longer
period systems or if their outbursts were simply not observed. We thus refrain from further analysis of these systems.

\begin{table*}
\centering
\begin{minipage}{5.5in}
\caption{Outburst properties of recurring outburst systems with known orbital periods.}
\label{tbl:recurringoutbursts}
\begin{tabular}{lccccccc}
\hline
System & Orbital & \# of Outbursts & Observation &Recurrence & Duration & Strength \\
& Per. (min) & Observed & Span (d) & Time (d) & (d) & (mag)\\
\hline
PTF1J1919+4815$^a$ & 22.5 & \nodata & \nodata & $36.8\pm0.4$ & $\mathord{\sim}13$ & 3\\
CR Boo$^b$ & 24.5 & \nodata$^c$ & 3445 & $47.6\pm4.8$ & $\mathord{\sim}24$ & 3.3\\
KL Dra$^a$ & 25.0 & \nodata & \nodata & 44--65 & $\mathord{\sim} 15$ & 4.2 \\
V803 Cen$^a$ & 26.6 & \nodata$^c$ & 2545 & 77 & \nodata & 4.6 \\
PTF1J0719+4858$^a$ & 26.8 & 23 & 2581 & 65--80 & $\mathord{\sim} 18$ & 3.5 \\
SDSSJ0926+3624 & 28.3 & 9 & 3462 & $160\pm20$ & $\mathord{\sim}20$ & 2.4 \\
CP Eri  & 28.7 & 13 & 2691 & $108\pm13$ & $\mathord{\sim}20$ & 4.2 \\
PTF1J0943+1029  & 30.4 & 10 & 3645 & $110\pm14$ & $\mathord{<}30$ & 4.1 \\
V406 Hya  & 33.8 & 5 & 2540 & $280\pm50$ & $\mathord{<}100$ & 5.9 \\
PTF1J0435+0029 & 34.3 & 2 & 2629 & $365\pm60$ & $\mathord{<}60$ & 5.1 \\
2QZJ1427-0123 & 36.6 & 3 & 3455 & $540\pm60$ & $\mathord{<}50$ & 4.3\\
\hline
\end{tabular}
\smallskip \\
Definitions of the properties shown here are in Section \ref{sec:outburst_def}.\\
$^a$ Properties presented here (except observation details) are from the literature. See Table \ref{tbl:amcvns} for references.\\
$^b$ The reported data are from only the second half of CR Boo observations presented in this paper (see Section \ref{sec:crboo}).\\
$^c$ We do not count the number of outbursts due to the complicated and rapidly changing nature of the light curve.\\
\end{minipage}
\end{table*}

\subsection{``Single Outburst'' Systems}
\label{sec:single}

\begin{figure*}
\includegraphics{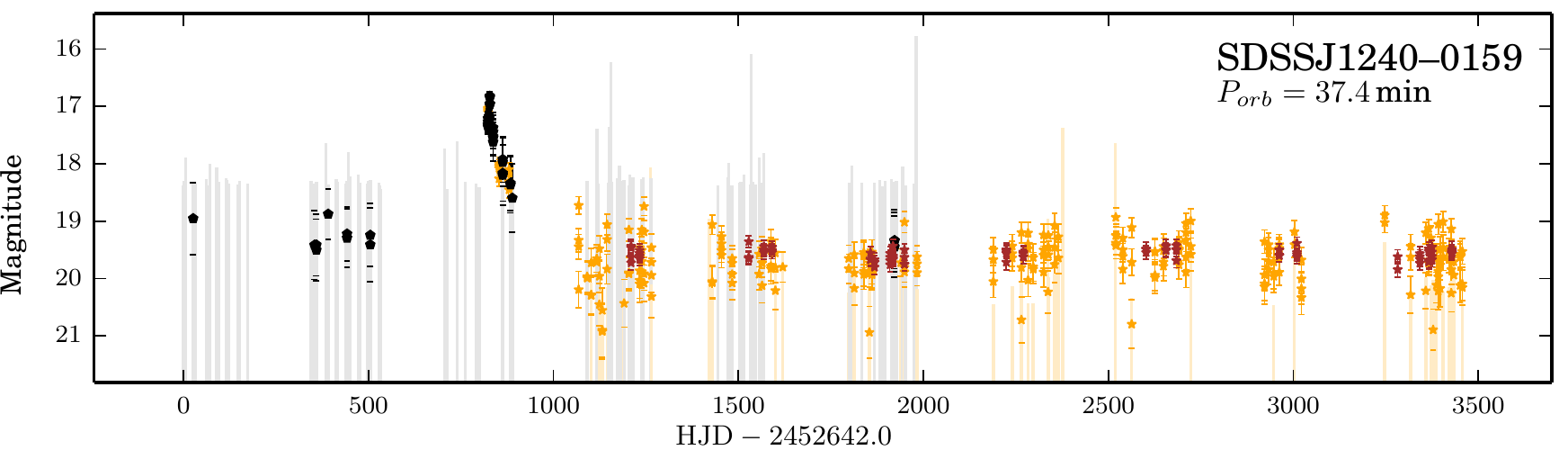}\\
\includegraphics{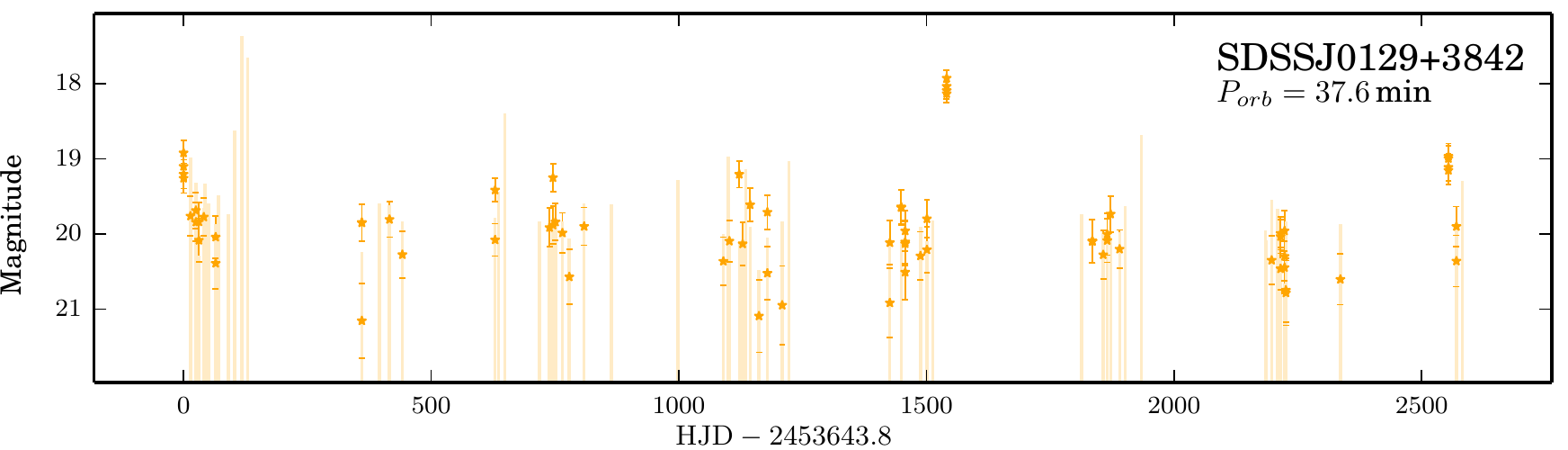}\\
\includegraphics{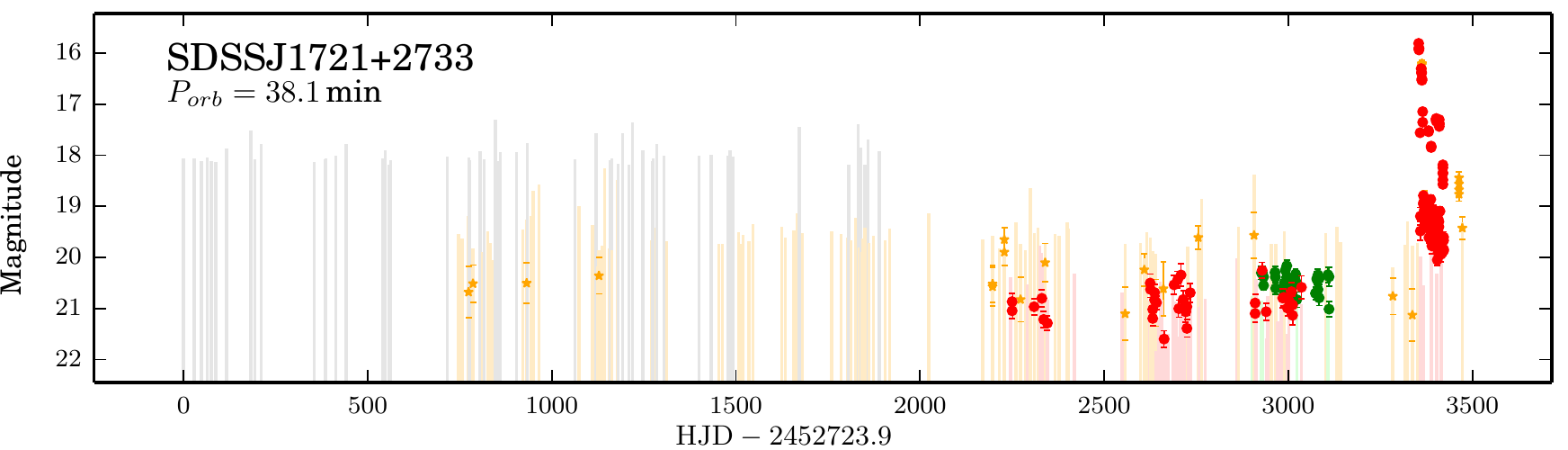}\\
\includegraphics{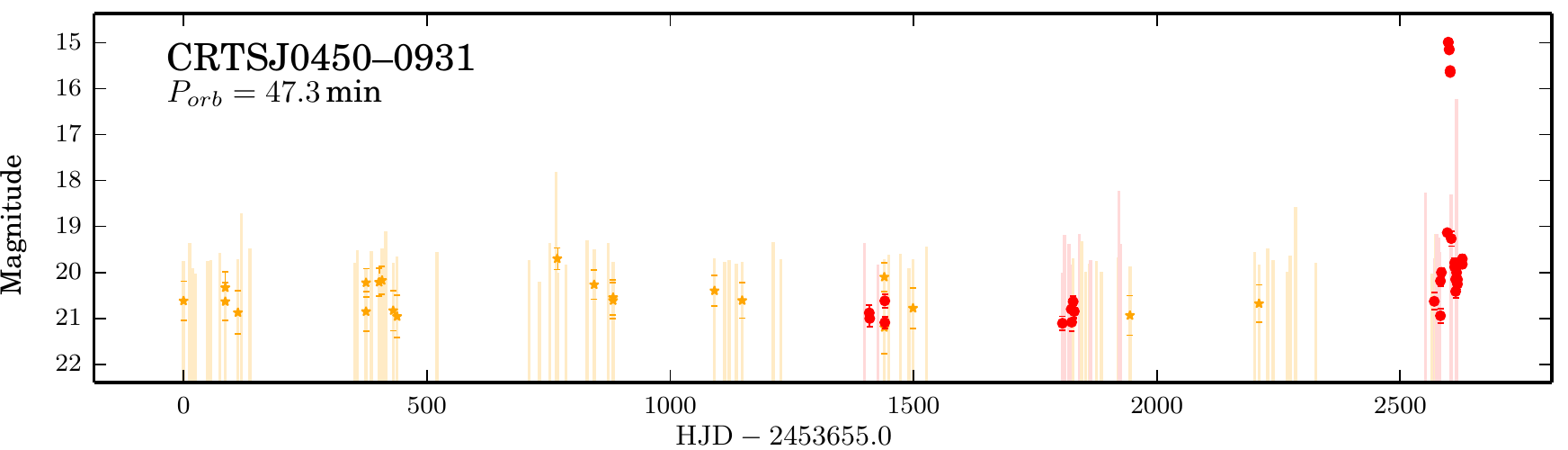}
\caption{Light curves of the four outbursting AM CVn systems with only one recorded outburst. All systems have longer
orbital periods than the regularly outbursting AM CVn systems. In the case of SDSSJ0129+3842, two additional possible
outbursts are visible, but they do not meet our criteria for an outburst (Section \ref{sec:outburst_def}). \newline
\textbf{Legend:} black = LINEAR; yellow = CSS; maroon = MLS; red = PTF $R$; green = PTF $g'$. The tops of the vertical lines (colour-coded to match the survey) are limiting magnitudes for non-detections.\label{fig:lcs:single1}}

\end{figure*}

\begin{figure*}
\includegraphics{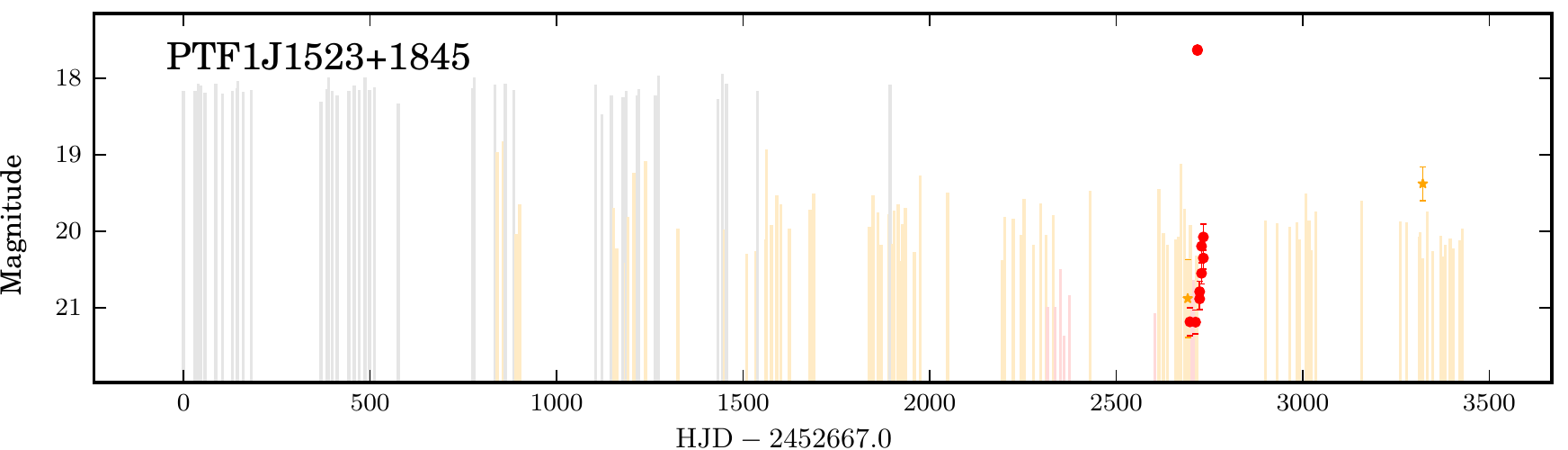}\\
\caption{Light curve of PTF1J1523+1845, a single outburst AM CVn system with no known orbital period.\newline
\textbf{Legend:} black = LINEAR; yellow = CSS; red = PTF $R$. The tops of the vertical lines (colour-coded to match the survey) are limiting magnitudes for non-detections.\label{fig:lcs:single2}}
\end{figure*}

\begin{figure}
\includegraphics{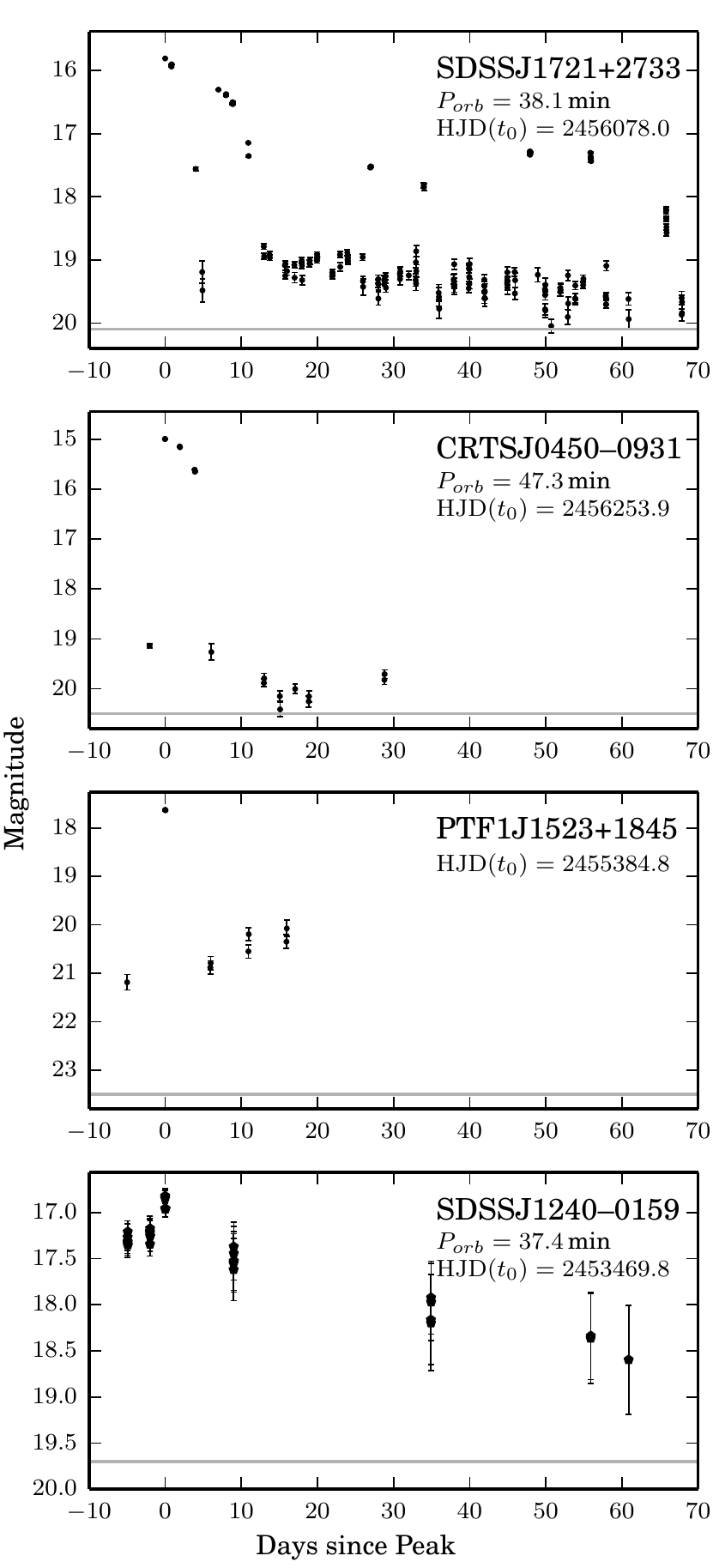}
\caption{A plot of the outburst light curves for four of the single outburst systems. SDSSJ1240--0159 is from LINEAR data and the
rest are PTF $R$-band data. The gray line indicates the quiescence level of each system. We note the similarity between the light
curves of SDSSJ1721+2733 and CRTSJ0450--0931, and, to a lesser extent, likely due to lack of data, PTF1J1523+1845. All three
systems show a sharp rise, a fall within 10 days, and a gradual decline towards quiescence. On the other hand, the light curve of
SDSSJ1240--0159 shows a gradual decline from peak and is still $\mathord{>}1.5\,$mag brighter than quiescence 60\,d
from the peak of the outburst.}
\label{fig:singleoutburstlcs}
\end{figure}

Seven of the known AM CVn systems have only had a single outburst recorded. 
We present the light curves of these systems in Figures \ref{fig:lcs:single1} and \ref{fig:lcs:single2}. Drawing on our observations, as well as
those reported in the literature, we list outburst times and strengths, as well as the probability of a missed outburst, in Table \ref{tbl:singleoutbursts}. 
We present the outburst light curves for four of the systems with the most details in Figure \ref{fig:singleoutburstlcs}.

\begin{table}
\centering
\begin{minipage}{\columnwidth}
\caption{Details of single outbursts.}
\label{tbl:singleoutbursts}
\begin{tabular}{lcccc}
\hline
System  & Outburst & Strength$^a$ & Probability of\\
& Date & (mag) & Missed Outburst \\
\hline
SDSSJ0129+3842 & 2009 Nov 29 &  $\mathord{\sim}5.4$ & $0.78\pm0.02$\\
CRTSJ0450--0931 & 2012 Nov 22 &  $\mathord{\sim}5$ & $0.75\pm0.02$\\
SDSSJ1240--0159 & 2005 Mar 15 &  $\mathord{\sim}6$ & $0.18\pm0.01$\\
PTF1J1523+1845 & 2010 Jul 07 &  $\mathord{\sim}5.8$ & $0.78\pm0.02$\\
SDSSJ1721+2733 & 2012 May 30 &  $\mathord{\sim}5$ & $0.59\pm0.02$\\
SDSSJ2047+0008 & 2006 Oct 12 &  $\mathord{\sim}5$ & $1.0$\\
\hline
\end{tabular}
\smallskip\\
The data presented in this table are drawn from a combination of the referenced papers and the light curves presented here.
Systems are arbitrarily ordered in terms of RA.\\ 
$^a$ The numbers presented here are lower bounds since the outburst peak was not always caught.\\
\end{minipage}
\end{table}

We focus on the data here, and leave out discussion of these systems and whether they are truly one-time outbursts until our discussion in Section \ref{sec:disc:single}.
The most important question to answer is to calculate the probability of a missed outburst.
We use a Monte Carlo approach where, for each of 1,000 iterations for each system, we tested whether an outburst starting at a random time
between the start and end points of the light curve would be detected. A system in outburst was assumed to be detected
if it was 1.5\,mag above quiescence and greater than the limiting magnitude for that exposure. We required at least two detections over
the course of the outburst. This itself was repeated 100 times, and the standard deviation of these 100 runs is the reported errors
for the probability of non-detection. The detection threshold was set in agreement with our definition of an outburst in
Section \ref{sec:outburst_def} and the scatter of points in quiescence for all these systems was $\mathord{\sim}0.5\,$mag.

For this to work effectively, we must use a reasonable model of the light curve. We note that for all but SDSSJ1240--0159, the post-peak
outburst light curve consists of a sharp decline that reaches 1--2\,mag above quiescence within 10\,d, and then a gradual decline over 
30-60\,d. We base this not only on our data (Figure \ref{fig:singleoutburstlcs}) but on similar light curves for SDSSJ0129+3842 in Figure 4 of R12 and SDSJ2047+0008 in Figure 4 of \citet{2008AJ....135.2108A}. We model all three systems by using
an inverse parabola that reaches 1.5\,mag above quiescence after 10\,d, and then a linear decline over the next 50\,d back to quiescence.
The only difference in our model between the systems is the initial outburst peak magnitude. In the case of
SDSSJ1240--0159, we assume a simple linear decline from peak to quiescence over 80 days. This difference accounts for the
significantly different shape of the outburst (Figure \ref{fig:singleoutburstlcs}).
The results of these calculations are listed in Table \ref{tbl:singleoutbursts}. 

We make three observations here based on these results. First, it is not surprising that SDSSJ2047+0008 was not detected in our data, 
given its short outburst duration and quiescent magnitude of $g'\sim24$ \citep{2008AJ....135.2108A}. 
Second, out of the rest of the systems, only SDSSJ1240--0159 is likely to have not had a missed outburst. Its outburst
shape, as noted earlier, is very different than the other systems. Finally, SDSSJ1721+2733 shows re-brightening events during its
decline (see Figure \ref{fig:singleoutburstlcs}), something also reported for SDSSJ0129+3842 \citep{2011arXiv1104.0107S}.

\subsection{Other Variability}
\label{sec:other}

\begin{figure*}
\includegraphics{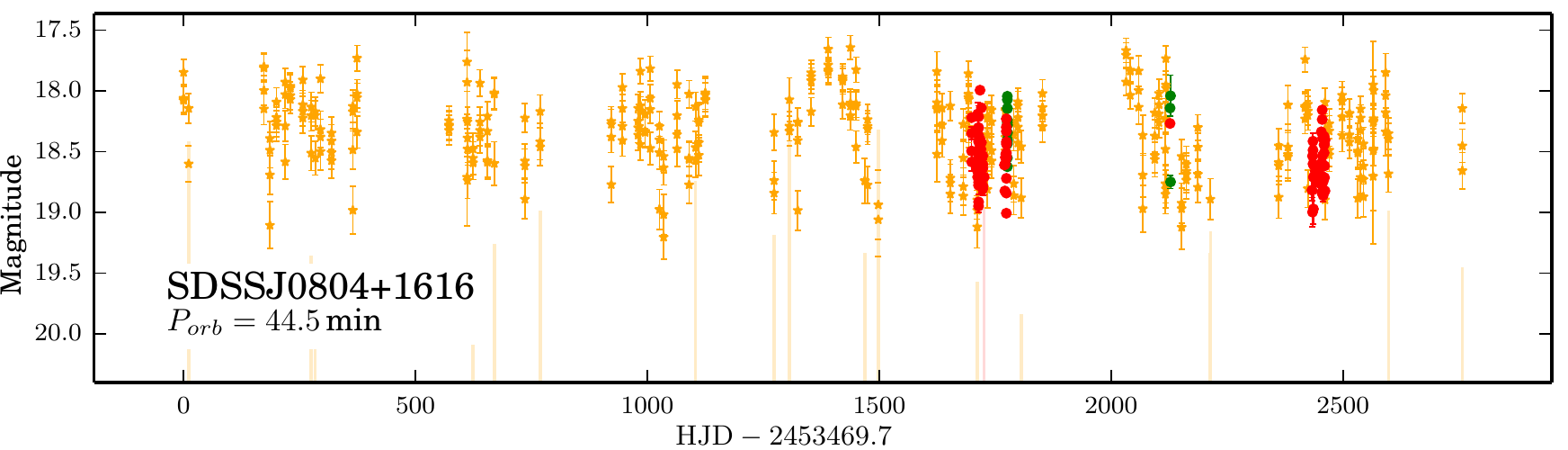}\\
\includegraphics{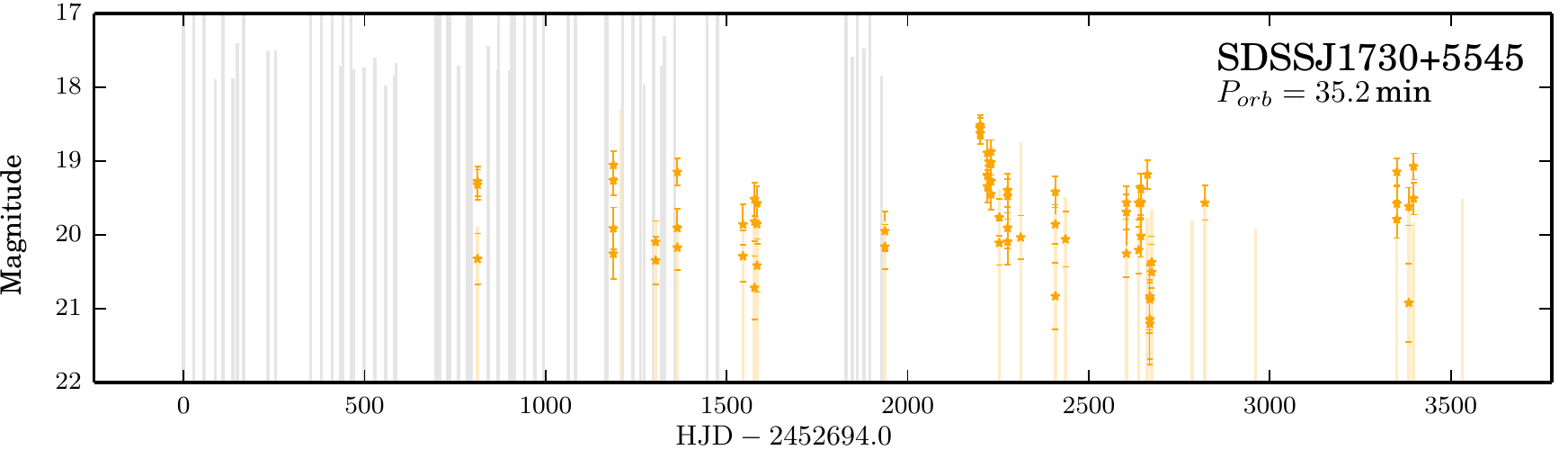}
\caption{Light curves of two systems with non-outburst variability (Section \ref{sec:other}). SDSSJ0804+1616 is possibly a magnetic system \citep{2009MNRAS.394..367R}
and shows non-periodic variability akin to that seen in magnetic CVs. SDSSJ1730+5545 shows a potential outburst, but one which does not
meet our criteria. \newline
\textbf{Legend:} black = LINEAR; yellow = CSS; red = PTF $R$; green = PTF $g'$. The tops of the vertical lines (colour-coded to match the survey) are limiting magnitudes for non-detections.\label{fig:lcs:unknown}}
\end{figure*}

R12 noticed that SDSSJ0804+1616 showed significant variability,
but not of the typical outburst variety. Instead, it showed irregular variability
with an amplitude of $\mathord{\sim}1\,$mag. The light curve we present in Figure \ref{fig:lcs:unknown}
confirms this variability over 7\,yrs.  We find no discernible period, although the time-scale of the variability
could be as short as 1--2 nights, based on several nights where the target was observed $\mathord{\sim}15$ times in one night by the PTF.
\citet{2009MNRAS.394..367R} suggested that SDSSJ0804+1616 may be a magnetic system. Similar light curves
have been observed in PTF for magnetic CVs \citep{MARGONIP}, strengthening the argument that SDSSJ0804+1616
is, in fact, a magnetic system.

We also present the light curve of SDSSJ1730+5545 in Figure \ref{fig:lcs:unknown}. The light curve contains what appears to be the tail end of an outburst. However,
despite multiple detections at $\mathord{\sim}1.5\,$mag brighter than the median magnitude, it fails to meet our criteria for the definition
of an outburst. Similarly, SDSSJ0129+3842 also shows at least two other candidate outbursts, both of which fail to meet our criteria.
We are reluctant to loosen the criteria, however, as SDSSJ1730+5545 is the only system where just a partial outburst may have been detected. We note that SDSSJ1730+5545's recently measured orbital period of 35.2\,min \citep{2014MNRAS.437.2894C}
places it at the long end of the outbursting orbital period regime, but poor coverage makes the concrete detection of an
outburst difficult given outburst recurrence times of systems with similar orbital periods.

\section{Results and Discussion}
\label{sec:discussion}

\subsection{AM CVn System Evolution}
\label{sec:evolution}

The composite light curves presented here allow us to see long-term changes in the photometric behavior of AM CVn systems. We
summarize the phenomenology of outbursting AM CVn systems in the following three stages of evolution:

\begin{enumerate}
\item When the mass transfer rate from the secondary ($\dot{M}$) falls below a critical value (believed to occur for $P_{orb} \gtrsim 20\,$min),
the accretion disc is no longer in a high state at all times and instabilities in the disc develop that lead to large amplitude photometric variations.
The light curves of the shortest-period systems in this study (CR Boo and V803 Cen) show that the transition from a stable high
state to ``regular'' outbursts
is in fact irregular with variations on long time-scales (years). The systems can spend most of their time in a high state with occasional excursions to the quiescent state (as has been observed exclusively for V803 Cen) or act as a more ``traditional'' outbursting system
--- remaining primarily in the quiescent state, with semi-regular outbursts to the high state. 

\item Only for $P_{orb}\gtrsim28\,$min do systems seem to settle into a more regular
pattern of quiescence with well-defined outbursts.
Between orbital periods of roughly 28\,min and 37\, min, AM CVn systems are primarily quiescent with somewhat regular outbursts,
the properties of which exhibit a gradual process of
a power law increase in recurrence time (see Section \ref{sec:recurtimevsperiod} for details).
Normal outbursts still occur, but are rarer and longer than in shorter-period systems. 

\item At longer orbital periods, $\dot{M}$ has decreased significantly and systems experience rare outbursts, if any. 
These systems may be the analogs to WZ Sge systems among the CVs, but the short outburst
durations ($\mathord{\sim}$10--15\,d) of all known systems except SDSSJ1240--0159 do not fit with this analogy.
One possible explanation is that such short outbursts are the equivalent of the normal outbursts
seen in much shorter-period systems (e.g., Section \ref{sec:cperi}). The outburst of SDSSJ1240--0159, which shows
a significantly longer duration than the remaining systems, would then be a super-outburst. Its outburst properties
are, in fact, consistent with the relations we find in Section \ref{sec:recurtimevsperiod}.
If this proposal is correct, then the recurrence time of these shorter-duration outbursts could be on the order of years,
while the recurrence time of super-outbursts could be decades.
Such a recurrence time would be consistent with
those seen in WZ Sge systems, but no normal outbursts have been observed in WZ Sge systems
\citep{2007MNRAS.375..105M}. However, the significantly different composition of the systems (He-rich vs. H-rich) and
the resulting significant difference in both separation between the components and component temperatures may
account for this difference in behavior. Additional study of the recently discovered He-rich CVs with orbital
periods similar to those of the longest known AM CVn systems \citep{Breedt:2012kx} may help resolve this question.
\end{enumerate}

It is obvious that orbital period is not the only factor influencing the behavior of these systems, and other factors, likely the component 
masses, donor composition, and donor entropy will play a role. For example, V406 Hya has significantly stronger outbursts than other 
systems of comparable orbital periods (see Table \ref{tbl:recurringoutbursts}). Additionally, transitions between states may
result in unstable photometric behavior: CR Boo and SDSSJ0926+3624 are possible examples of such systems.

\subsection{Outburst Behavior vs. Orbital Period}
\label{sec:recurtimevsperiod}

The change in outburst behavior with orbital period appears to be gradual, rather than abrupt.
While there are only data for a limited number of systems, these are enough to find an approximate relation.
For the outburst recurrence time and duration
we chose to use a power law model, while for the strength, $\Delta$mag, we used a linear model  in magnitudes
(this corresponds to an exponential model in flux). These choices are somewhat arbitrary and are only a simple phenomenological 
approximation to any physical relation. An exponential model fits the outburst recurrence time and duration equally well
(see Appendix \ref{app:fit}) but a power law is consistent with the orbital evolution equations proposed for
AM CVn systems \citep{Faulkner:1972cr}. Using the values from Table \ref{tbl:recurringoutbursts}, we find the following relations,
\begin{figure}
\includegraphics{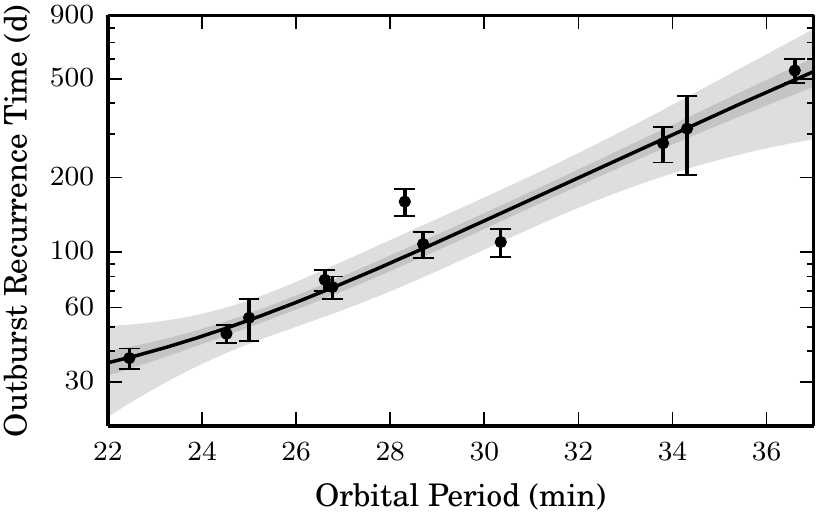}\\
\includegraphics{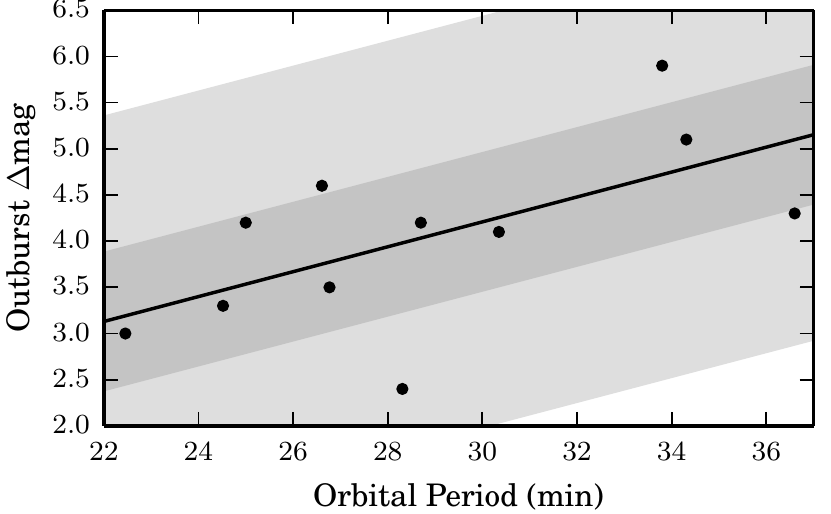}\\
\includegraphics{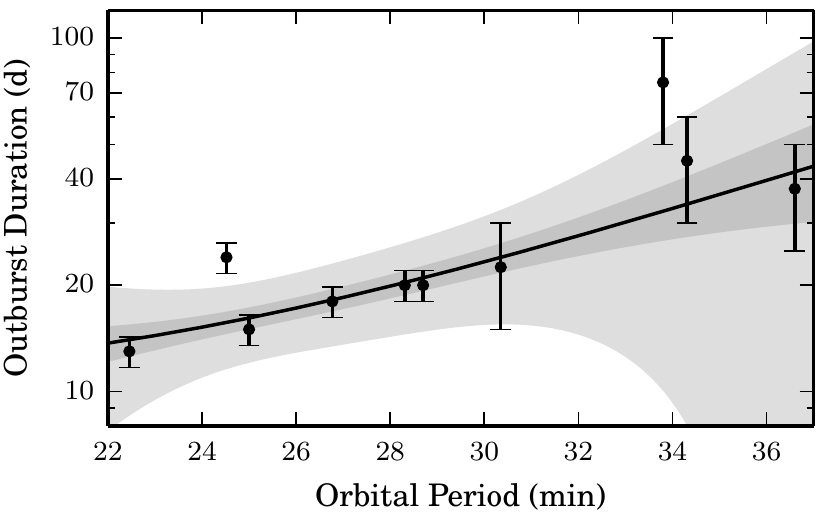}\\
\caption{Plots of outburst properties vs. orbital period.
The solid line is a best-fitting model.
For the recurrence time and outburst duration we used an power law model, while for the outburst magnitude we used a linear model
(which corresponds to an exponential model in flux). The former are plotted using a logarithmic scale on the y-axis, while we
note that magnitudes are already logarithmic.
The darker-shaded areas represent
the $1\sigma$ errors while the lighter-shaded areas represent the $3\sigma$ errors. For the recurrence time and duration, we use the fit errors.
For the outburst strength, we use the standard deviation of the residuals. Full details of the fits are given in Appendix \ref{app:fit}.}
\label{fig:outburstrelation}
\end{figure}
\begin{align*}
P_{recur}  & = (1.53\times 10^{-9}) P_{orb}^{7.35} + 24.7 \\
\Delta\mathrm{mag} &= 0.13 P_{orb} - 0.16 \\
t_{dur} &= (2.53\times10^{-6}) P_{orb}^{4.54} +10.6,
\end{align*}
where $P_{orb}$ is the orbital period in minutes, $P_{recur}$ is the outburst recurrence time in days, $\Delta$mag is the
strength of the outburst, and $t_{dur}$ is the duration of the outburst in days. A plot of these
quantities, together with the best fits, are shown in Figure \ref{fig:outburstrelation}. We provide complete fit details, including
information about the fit errors, in Appendix \ref{app:fit}. The outburst recurrence time is a much better fit
than the duration or strength --- this may be due to either measurement errors or because AM CVn systems
vary more in outburst strength and duration than in recurrence time. We also do not account for the progenitor
type of each system \citep[e.g., ][]{2010MNRAS.401.1347N}, although it is possible that this has an impact on system outburst behavior.

Verification of these relations will require significant additional period measurements. We note that these relations do not
apply to systems with only one observed outburst,
and we do not recommend applying them to systems with only a few observed outbursts. It is highly likely that,
particularly at the long-period end, these relations are not accurate due to the lack of data in that period regime.
In particular, the single outburst systems identified in this paper typically show an outburst duration of only 10--15\,d (see Figure
\ref{fig:singleoutburstlcs}), whereas $t_{dur}$ trends towards 50\,d at a similar orbital period.

We also note that these relations apply only to optical wavelengths. AM CVn systems have been poorly studied
in other wavelengths, although the few systems observed have been seen to vary in other wavelengths.
In particular, KL Dra was observed in UV and X-ray by \citet{Ramsay:2010uq} and SDSS1043+5632 shows variability
of 3.7\,mag over 26 NUV observations in the Second GALEX Ultraviolet Variability Catalog \citep{2008AJ....136..259W}.
Future UV missions (e.g., ULTRASAT; \citealt{2014AJ....147...79S}) will help better explain UV variability in AM CVn systems.

\subsubsection{Are the outburst property relationships realistic?}

Our fits make two substantial assumptions: first, that all three properties we model (outburst recurrence time, duration, and strength)
are increasing with respect to $P_{orb}$ and,
second, that the relationships are dependent only on $P_{orb}$. We aim to verify whether both of these are true.

To ascertain the correlation between $P_{orb}$ and the outburst properties, we calculate the Spearman rank
correlation coefficient for 100,000 unique, random permutations of the property values. The fraction of the
permutations for which the coefficient is greater than for the data points in order (indicating that this set of points
is more correlated with $P_{orb}$ than the original set) is the p-value that the property
is not correlated with the orbital period (with a p-value of 0 indicating high probability of correlation and 1 indicating a
low probability of correlation). We find that the p-values for recurrence time, outburst strength,
and duration are, respectively, 0.00, 0.026, and 0.0057. These indicate that the recurrence time and, to a slightly
lesser extent, the duration, are strongly correlated with $P_{orb}$, while outburst strength is slightly less correlated.

We thus conclude that it is very likely that all the properties are correlated with $P_{orb}$. However, are they only dependent on
$P_{orb}$ or do other system properties influence this as well?
 If, in fact, the relationships are dependent only on $P_{orb}$ and the correct model is being used, then the residuals
should be distributed around zero with a normal distribution. We use the Shapiro-Wilk test \citep{SHAPIRO01121965} to
calculate the probability that the residuals are taken from a normal distribution (although we find that related tests provide
similar results). The p-values for the recurrence time, outburst strength, and outburst
duration are, respectively, 0.03, 0.81, and $1.1\times10^{-4}$; these represent the probability of observing such residuals
had they been normally distributed.

While the outburst strength is normally distributed, both the recurrence time and outburst duration are likely not.
This indicates that the models for these two properties are too simplistic. However, given the lack of additional
data for systems (e.g., component masses), these are likely the best approximations that can be determined at the present time.

\subsubsection{Prediction of Orbital Periods}
\label{sec:prediction}

The measurement of AM CVn system orbital periods is a difficult process, particularly for the faint systems discovered recently.
The relation between orbital period and outburst recurrence time presented in Section \ref{sec:recurtimevsperiod}
allows us to estimate periods for systems not yet measured. Four systems show multiple outbursts with a consistent recurrence
time and have unknown orbital periods. We provide estimated orbital periods for them, along with their outburst properties in
Table \ref{tbl:unknownrecur}. We caution that these are estimates to serve primarily in observation planning. Errors are derived
from a combination of fit parameter errors and outburst recurrence time errors.

\begin{table*}
\centering
\begin{minipage}{5.5in}
\caption{Outburst properties of recurring outburst systems with unknown orbital periods.}
\label{tbl:unknownrecur}
\begin{tabular}{lccccccc}
\hline
System & \# of Outbursts & Observation &Recurrence & Duration & Strength & Est. Orbital \\
 & Observed & Span (d) & Time (d) & (d) & (mag) & Per. (min)\\
\hline
PTF1J2219+3135 & 9 & 2726 & $64\pm5$ & $\mathord{<}26$ & 4.4 & $26.1\pm0.74$\\
SDSS1043+5632 & 9 & 3477 & $99\pm12$ & $\mathord{<}55$ & 3.4 & $28.5\pm0.92$\\
PTF1J1632+3511 & 3 & 3541 & $230\pm35$ & $\mathord{<}80$ & 5.2 & $32.7\pm1.1$\\
CRTSJ0744+3254 & 12 & 3100 & $239\pm36$ & $\mathord{<}65$ & 3.8 & $32.9\pm1.1$\\
\hline
\end{tabular}
\smallskip \\
Definitions of the properties shown here are in Section \ref{sec:outburst_def}. The estimated orbital periods
are based on outburst properties and their calculation and accuracy are described in Section \ref{sec:prediction}.
Errors are derived from a combination of the outburst recurrence time error and the fit error.
The outburst duration times for all of these systems are upper bounds due to lack of data to
find a better estimate.
\end{minipage}
\end{table*}

\subsubsection{Single Outburst Systems}
\label{sec:disc:single}

In Section \ref{sec:data}, we separated the outbursting AM CVn systems into those that showed regular outbursts, and those for
which only a single outburst has been observed.
We also showed in Table \ref{tbl:singleoutbursts} that it is highly likely that we missed an outburst for most of the
systems. Only for one system did we find a probability of a missed outburst below 50\%, while four out of
six have missed-outburst probabilities of $\mathord{\geq}75\%$.

Before the discovery of a 47\,min photometric period in CRTSJ0450--0931 \citep{Woudt:2013kx},
only systems with $P_{orb}$ below 40\, min were believed to outburst. Even more recently,
an outburst in SDSSJ0902+3819 --- a system with a spectroscopically measured orbital period
of 48.3\,min \citep{2010ApJ...708..456R} --- was observed in outburst by \citet{2014arXiv1407.4196K}.
While it is not known if all systems with similar orbital periods experience outbursts, the discovery
of two systems indicates this is likely not a unique phenomenon.

Using the relation in Section \ref{sec:recurtimevsperiod}, the recurrence time for a $38\,$min system is 2\,yrs.
The recurrence time of a $48\,$min system according to our relation is 9.6\,yrs. If we assume our relation holds
at such a long orbital period, then even the data presented here does not extend far enough back to contain
even two outbursts. The relatively short nature of these outbursts and the faintness of many of the systems
makes such detections even more difficult. Only three singe-outburst systems were detected in outburst in the
PTF data, four systems were detected in 7 yr of CSS data, and one system was detected in 5.5 yr of LINEAR data.
For these reasons, we believe that most of the ``single'' outburst systems follow the same principles as
shorter-period orbital systems, but, given
their short outburst duration (see Sections \ref{sec:single} and \ref{sec:evolution}), long recurrence times, and
faint quiescent magnitudes, are simply difficult to detect in outburst. 

\subsection{Implications for Discovery of AM CVn Systems}

The relationships between orbital period and outburst properties developed in Section \ref{sec:recurtimevsperiod} allow us to
calculate the detection probability, $\text{p}(P_{orb}, m_{qui})$, of an outbursting AM CVn system by a synoptic survey with a known cadence and limiting magnitude.
We can use these results to estimate the number of outbursting AM CVn systems with $20\,\text{min}<P_{orb}<37\,$min
that a survey could discover.
Such a calculation involves two elements. First, we must find the detection probability of an AM CVn system that has
a specific orbital period and quiescent magnitude. Second, we need a model for the Galactic distribution of AM CVn systems.
Here, we calculate the number of systems that could be discovered by two model surveys based on the CSS and the PTF.

\subsubsection{Survey Definition and System Detection Probability}
We begin by defining our surveys. 
We assume no weather interruptions, and  normal-distributed limiting
magnitudes with $\sigma=0.5\,$mag around the median limiting magnitude of the survey. We do not account here for crowding
and assume perfect detections (e.g., no artefacts). 
For the CSS-like survey, we assume four exposures per night over 30\, min, taken every 2 weeks \citep{2009ApJ...696..870D},
and a median magnitude of $V=19.5$. We assume that each field is observed for $\mathord{\sim}200\,\text{d}=15$ observations per year,
for 7 years.
For the PTF-like survey, we assume 2 exposures per night over 1\,h, but with a cadence of 4\,d and a limiting magnitude of $V=20.5$.
We assume that each field is observed for $\mathord{\sim}3$ months (20 observations) for 3 years.
Lastly, we assume that both surveys cover Galactic latitudes of $15<b<90$
at all Galactic longitudes.

We now construct an outburst light curve model. Although we constructed such a model
for the calculation of non-detection probabilities in Section \ref{sec:single}, that model was only applicable to systems with
$P_{orb}>37\,$min. The light curve profile (see Section \ref{sec:regular} of this paper and Figure 4 of R12) of
outbursting systems with $P_{orb}<37\,$min is substantially different. Thus, we model the outburst as a sudden rise to
the outburst magnitude ($\Delta\text{mag}+m_{qui}$, as defined in Section \ref{sec:recurtimevsperiod}), and a gradual decline
over $t_{dur}$ days to 0.5\,mag above $m_{qui}$, with a return to quiescence thereafter.

To calculate the probability, $\text{p}(P_{orb}, m_{qui})$, we use a Monte Carlo approach. For every $P_{orb}$ and $m_{qui}$,
we calculate the light curve at the simulated exposure times using a random start time for the outburst sequence. We determine
whether a particular light curve was detected based on the criteria in Section \ref{sec:outburst_def}. Briefly, we required at least 2
consecutive detections (defined as being brighter than the limiting magnitude) within 15\,d that were $\mathord{\geq}0.5\,$mag above the quiescence magnitude. We note that we only use the 0.5\,mag
above quiescence criterion here, as opposed to the $3\sigma$ criterion. However, the error of observations at the $5\sigma$ limiting magnitude
should be $\mathord{\sim}0.2\,$mag, which is consistent with these criteria here. We caution that these criteria for outbursts, and the ones
generally applied in this paper, are designed only to ignore fake outbursts. In a real survey, one would also want
to select against short outburst-like events, such as M-dwarf flares.  We simulate 1,000 systems for each
$P_{orb}$ and $m_{qui}$. We repeat this process 500 times, and take the mean and standard deviation of the number of systems detected
over the number of systems simulated as the detection probability and its associated error.
We calculate the detection probability for $20\,\text{min}\leq P_{orb}\leq37\,$min in 0.2\,min steps and for $17\leq m_{qui}\leq26$
in 0.2\,mag steps, and interpolate for intermediate values.

\begin{figure}
\includegraphics{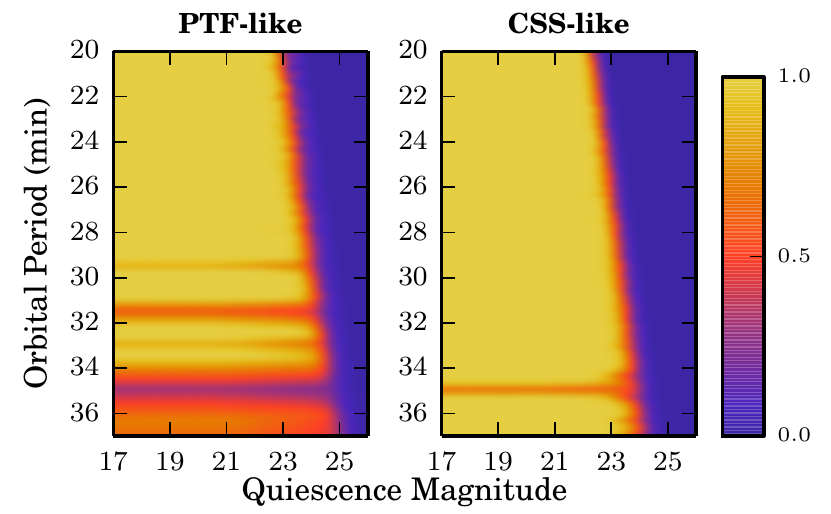}
\caption{A plot of the detection efficiency of AM CVn systems given an orbital period and quiescent magnitude. A significant decrease
in efficiency is seen at $P_{orb}=35.8\,$min, as at this time the recurrence time is about one year. The PTF survey goes slightly deeper,
but this is not as large an effect due to the longer baseline of the CSS. The PTF suffers at longer-orbital periods as the recurrence times
increase to several years. \label{fig:surveyefficiency}}
\end{figure}

In Figure \ref{fig:surveyefficiency} we show the detection efficiency of our surveys given $P_{orb}$ and $m_{qui}$.
We caution that these models do not account for weather
and other scheduling irregularities and, particularly in the case of the PTF-like survey, are only vaguely similar to the cadence of the
survey they emulate. As expected, longer-period systems can be detected to fainter
magnitudes given their increased strength, but are not as
well-detected by the PTF-like survey due to its shorter baseline, relative to the $\mathord{>}1\,$yr
recurrence times at these orbital periods.
The PTF-like survey is able to detect slightly fainter systems due to being deeper, but the longer baseline of the CSS-like survey
removes this advantage.

\subsubsection{System Evolution Models}
\label{sec:sysevol}

Now that we have $\text{p}(P_{orb}, m_{qui})$, we must model the population of AM CVn systems.
First, we find the fraction of AM CVn systems at each orbital period.
The orbital evolution of AM CVn systems is believed to involve only the effects of gravitational
wave radiation and mass transfer (\citealt{1967AcA....17..287P}; but see \citealt{2005ApJ...624..934D}
for a more complex evolutionary model). We assume that the
percentage of systems at a given $P_{orb}$ is equal to the amount of time the system
spends at that orbital period over the lifespan of the system, which we define to be from $P_{orb}=5\,$min to 80\,min.
This ignores any changes in the birth rates of these systems.

We evolve a system
with $M_{acc}=0.6M_\odot$ and $M_{don}=0.25M_\odot$ from $P_{orb}=5\,$min to longer orbital periods.
The masses are arbitrary, but are in agreement with models and with the measured masses of
the components of SDSSJ0926+3624 \citep{2011MNRAS.410.1113C}. To simplify the calculations,
we fit these results with a power law and take its derivative, such that $f_{syst}$ is the
  fractional number of systems per orbital period bin and $P_{orb}$ is in minutes
\begin{equation}
f_{syst}(P_{orb})=(4.0\times10^{-7}) {P_{orb}}^{2.66} . \label{eq:fsyst}
\end{equation}
This equation is normalized, such that
$\int_{5\,\text{min}}^{80\,\text{min}} f_{syst} (P_{orb})\, \text{d} P_{orb}=1$ and hence
the integral of $f_{syst}$ between two orbital periods will yield the fraction of a system's lifetime spent between those
orbital periods and thus the fraction of known AM CVn systems we expect to observe between the orbital periods.
We note that an analytic derivation of the orbital period derivative, $\dot{P}_{orb}(P_{orb})$, in the limit $M_{don}\ll M_{acc}$
yields $\dot{P}(P_{orb})\propto P_{orb}^{-8/3}$, consistent with the numerical fit.

We use the same Galactic population distribution model as \citet{2001A&A...368..939N},
\begin{equation}
\rho(P_{orb}, R,z)=\rho_0 f_{syst}(P_{orb}) e^{-R/H}\text{sech}(z/h)^2\,\text{pc}^{-3}, \label{eq:rho}
\end{equation}
where $R$ is the radius from the centre of the Galaxy, $z$ is the distance above the Galactic plane, $\rho_0$ is the population density
at the centre of the Galaxy, $H$ is the scale distance, and $h$ is the scale height. We adopt, for the purposes of this calculation,
the same scale height (300\,pc) and scale distance (2.5\,kpc) as \citet{2007MNRAS.382..685R}.

The number of systems with orbital period $P_{orb}$ at a point $(r, b, l)$ when viewed from Earth can then be defined as
\begin{equation}
N_{obs}(P_{orb}, r, b, l)=r^2 \cos (b) \rho(P_{orb}, R,z) \text{p}(P_{orb}, m_{qui}), \label{eq:nobs}
\end{equation}
where $b$ is the Galactic latitude, $l$ is the Galactic longitude, and we can express $R$ in terms of $r$, $b$, and $l$ as
$\sqrt{r^2 \cos^2 b + R_{GC}^2+2 r \cos b \cos l}$. $R_{GC}$ is 8125\,pc, the distance from Sun to the Galactic Centre.

We calculate $m_{qui}$ using the distance, $r$, and
the same parameterization for the absolute magnitude as \citet{2007MNRAS.382..685R},
\begin{equation}
M_{qui}(P_{orb})=10.5+0.075(P_{orb}-30\,\text{min}), \label{eq:mqui}
\end{equation}
which is based on Figure 2 of \citet{2006ApJ...640..466B}. This value for the absolute magnitude is only based on the
temperature of the accretor and does not account for any luminosity from the disc. However, the disc has been measured
to account for only 30\% of an AM CVn system's luminosity \citep{2011MNRAS.410.1113C}, so this assumption should
provide a reasonable estimate.

\subsubsection{Simulated Survey Results}

We now combine our model for the detection efficiencies with that for the Galactic distribution to find the number of expected systems
with $20\,\text{min}\leq P_{orb}\leq37\,$min
that would be detected by our CSS-like survey and our PTF-like survey. 
We use the most recent published population density estimate for AM CVn systems from \citet{Carter:2013fk}, hereafter C13.
Since C13 gives the local population density as opposed to the density at the Galactic centre (where we defined $\rho_0$),
we set $\rho_0=(5\pm3)\times10^{-7}\,\text{systems}\,\text{pc}^{-3} e^{R_\odot/h}$.

We find that over the survey lifetime, our CSS-like survey would detect $(1.76\pm1.1)\times10^{-3}\,\mathrm{systems}\deg^{-2}$ or,
assuming a total coverage of $\mathord{\sim}20,000\deg^2$, a total of $35\pm21$ systems in total.
For our PTF-like survey, we find that it would detect $(1.52\pm0.91)\times10^{-3}\,\mathrm{systems}\deg^{-2}$
over the survey lifetime. With a coverage
of $\mathord{\sim}16,000\deg^2$, we would expect a total of $24\pm15\,$systems. Errors provided are only
based on the error provided for the population density estimate.

Have the CSS and the PTF detected as many systems as we would expect if the population densities
from \citet{Carter:2013fk} are correct? 
The CSS has detected 8 AM CVn systems in outburst with $20\,\mathrm{min}<P_{orb}<37\,$min, and another likely
four systems with orbital periods in this range. The PTF has detected 6 outbursting AM CVn systems in this orbital period range,
and an additional 3 systems with orbital periods likely to be in this range. This indicates that the surveys have detected,
respectably, only 34\% and 38\% of the estimated total, albeit with significant errors in these numbers.
This likely shows the value of a dedicated, systematic search for these systems, particularly given the recent results from
the partially-completed spectroscopic survey of all identified CRTS CVs \citep{2014MNRAS.443.3174B}.

We caution that our simulations did not account for several factors.
First, we did not account for scheduling irregularities
and we assumed a perfect cadence. PTF, in particular, uses variable cadences. A more realistic
study of PTF's AM CVn system detection efficiency based on the actual times of exposures is outside the scope of this paper.
An additional observational constraint is the
difficulty in confirming faint candidates. Systems with quiescent magnitudes significantly fainter than $g'\sim21$ cannot be
spectroscopically confirmed even with 8--10\,m class telescopes unless caught in outburst. These factors indicate
that while the CSS and the PTF likely contain additional systems, many may be faint and confirming these
systems will be extremely difficult.

Although this simulation considers regularly outbursting systems, we also need to consider
the probability of detecting longer-period systems.  If, in fact, longer-period systems do outburst as
we discuss in Section \ref{sec:disc:single}, and
the relation in Section \ref{sec:recurtimevsperiod}
(or a similar one) holds even for longer-period systems,
this implies that systems with orbital periods similar to CRTSJ0450--0931 and SDSSJ0902+3918
outburst on the decade time-scale. Such a time-scale is not
unreasonable, given the behavior of WZ Sge-type systems.
The majority of AM CVn systems are believed to be long-period systems
\citep{2001A&A...368..939N,Nissanke:2012fk} and faint. Specifically, we can approximate that
there are $\mathord{\sim}2.2\times$ more AM CVn systems with $37\,\mathrm{min}<P_{orb}<50\,$min
than with $20\,\mathrm{min}<P_{orb}<37\,$min using our evolutionary model.
Yet even if they are bright enough to be visible, only some will outburst during even a
decade-long synoptic survey (depending on the actual outburst recurrence time), and
of that sample, likely up to 75\% (Table \ref{tbl:singleoutbursts}) will be undetected
due to their short outbursts and the relatively sparse coverage of current
synoptic surveys. 

\subsection{Mass Transfer Rate vs. Outburst Recurrence Time}

The simplest hypothesis for the increase in outburst recurrence time with increasing orbital period would be that the
critical mass needed for  the disk instability, $M_{outburst}$, simply takes longer to accumulate as the mass transfer rate
decreases. Combining  our mass transfer rate model of an AM CVn system developed in Section \ref{sec:sysevol} with our
observed relationship  for the outburst  recurrence time, $t_{recur}$, in Section \ref{sec:recurtimevsperiod}, 
we find that $M_{outburst}\approx 10^{-10} M_\odot$ across the large orbital period range.  This lends strong support to the hypothesis 
that the outburst recurrence time is linked to $\dot{M}$. A more sophisticated model could be constructed using 
\citet{2005ApJ...624..934D} for $\dot{M}$ and the theory of \citet{2012A&A...544A..13K}  for $M_{outburst}$. 

\section{Conclusions and Further Work}
\label{sec:summary}

We have presented light curves of outbursting AM CVn systems drawn from three wide-area synoptic surveys, identified outburst 
recurrence times for all known outbursting systems with more than one observed outburst, and found relationships between the
orbital period and outburst strength, recurrence time, and duration. In particular, the light curve properties and resulting
relationships provide a powerful tool for understanding these and future systems. Although the current DIM model has been
successful in replicating normal and super outburst light curves of AM CVn systems \citep{2012A&A...544A..13K}, the much
more diverse behavior shown here (particularly the sudden changes in outburst behavior of CR Boo and SDSSJ0926+3624)
provides additional challenges to the model. The broader set of observations provided by synoptic surveys will hopefully
allow for the better development of AM CVn system outburst models. Lastly, we note that the approach
taken here is essentially qualitative in nature. A more rigorous, statistical approach to identifying and measuring outbursts
would provide significantly better results, particularly for those systems with few outbursts and/or few observations.

Our attempt to quantify the efficiencies of two different survey models will allow for a better understanding of the potential of
future synoptic surveys to identify new AM CVn systems. Confirmation
of these efficiencies with data on the actual observation schedules of the surveys will allow a better prediction of how many
systems remain in both surveys.

\section*{Acknowledgements}

We thank Lars Bildsten for helpful suggestions related to the recurrence time-orbital period relations. P. G. and T. P. thank the Aspen Center for Physics and the NSF Grant \#1066293 for hospitality during the preparation of this manuscript. E. O. O. is incumbent of
the Arye Dissentshik career development chair and
is grateful to support by
grants from the 
Willner Family Leadership Institute
Ilan Gluzman (Secaucus NJ),
Israeli Ministry of Science,
Israel Science Foundation,
Minerva, Weizmann-UK and
the I-CORE Program of the Planning
and Budgeting Committee and The Israel Science Foundation.

Observations obtained with the Samuel Oschin Telescope at the Palomar
Observatory as part of the Palomar Transient Factory project, a scientific
collaboration between the California Institute of Technology, Columbia
University, Las Cumbres Observatory, the Lawrence Berkeley National
Laboratory, the National Energy Research Scientific Computing Center,
the University of Oxford, and the Weizmann Institute of Science.
The CSS survey is funded by the National Aeronautics and Space
Administration under Grant No. NNG05GF22G issued through the Science
Mission Directorate Near-Earth Objects Observations Program.  The CRTS
survey is supported by the U.S.~National Science Foundation under
grants AST-0909182. The LINEAR program is sponsored by the National
Aeronautics and Space Administration (NRA No.~NNH09ZDA001N, 09-NEOO09-0010)
and the United States Air Force under Air Force Contract FA8721-05-C-0002.
This research has made use of NASA's Astrophysics Data System.

\bibliographystyle{mn2e}
\bibliography{amcvns}

\appendix
\section{System Model Fitting Details}
\label{app:fit}

\subsection{Outburst Recurrence Time and Duration Fits}
\label{app:exp}
In Section \ref{sec:recurtimevsperiod} and Figure \ref{fig:outburstrelation}, we provided
relationships between system properties and orbital period based on a fit to a model. In
particular, we fitted the known values for outburst recurrence time and outburst duration
time to a power law model (outburst $\Delta$mag was fitted to a linear model and is described in
Appendix \ref{app:lin}),
\begin{equation*}
y=\alpha P_{orb}^{\beta}+\gamma,
\end{equation*}
where $y$ is the property of the outburst, $P_{orb}$ is the orbital period in minutes, and
$\alpha$, $\beta$, and $\gamma$ are fit parameters.

To fit our observed values for the outburst properties, we used the \noun{NonlinearModelFit}
function in Wolfram Research's Mathematica 9.0.
The errors used for recurrence times are given in Table \ref{tbl:recurringoutbursts}.
For outburst duration, we assumed a 10\% error for all systems that do not have an upper limit in Table \ref{tbl:recurringoutbursts}.
For those with upper limits, we assumed that the duration was 75\% of the upper limit, with a 25\% error. These choices are
somewhat arbitrary, but are reasonable given the light curves.

\paragraph*{Recurrence Time Fit Errors.} 
The best-fitting values for the recurrence time are $\alpha=1.53\times10^{-9}$, $\beta=7.35$, and $\gamma=24.7$.
The elements of covariance matrix for the parameters are $\Sigma_{\alpha\alpha}=1.62\times10^{-17}$, 
$\Sigma_{\beta\beta}=0.567$, $\Sigma_{\gamma\gamma}=39.5$, $\Sigma_{\alpha\beta}=-3.03\times10^{-9}$,
$\Sigma_{\alpha\gamma}=-2.22\times10^{-8}$, and $\Sigma_{\beta\gamma}=4.09$. 

\paragraph*{Outdurst Duration Fit Errors.} 
The best-fitting values for the recurrence time are $\alpha=0.390$, $\beta=0.122$, and $\gamma=7.90$.
The elements of covariance matrix for the parameters are $\Sigma_{\alpha\alpha}=4.80\times10^{-10}$, 
$\Sigma_{\beta\beta}=5.99$, $\Sigma_{\gamma\gamma}=19.5$, $\Sigma_{\alpha\beta}=-5.36\times10^{-5}$,
$\Sigma_{\alpha\gamma}=-9.08\times10^{-5}$, and $\Sigma_{\beta\gamma}=10.1$. 

\paragraph*{Model Choice.}
We fit both the power law model described here and an exponential model. We then calculated the F-ratio
(defined as the ratio of the sum of the squares of the residuals) and found the p-value for this ratio. Additionally,
we performed the $\chi^2$ test on each model. We caution that both of these tests may provide erroneous values given the small
number of samples.

For the recurrence time, the exponential model was favored with a p-value of 0.4. The $\chi^2$ value for this model was also
slightly lower than for the power law model (53.5 vs. 56.2). Both of these differences are small. For the duration, neither
model was favored by the F-ratio and there was a difference of 0.7 in the $\chi^2$ values. Given the results from this test,
the possible impact of the small number of samples, and our preference for a power law model given the orbital evolution
equation, we chose the power law model.

\subsection{Outburst Strength Fit}
\label{app:lin}
For the outburst strength, we fit a linear model,
\begin{equation*}
\Delta\text{mag}=\alpha P_{orb}+\beta,
\end{equation*}
using the \noun{LinearModelFit} function in Wolfram Research's Mathematica 9.0.
We did not provide any weights as there is no good method of obtaining meaningful estimates of our errors.
We found a best fit of $\alpha=0.13$ and $\beta=0.16$. Errors are estimated calculating the standard deviation
of the residuals from the model, with $\sigma=0.74\,$mag.

\end{document}